\documentclass[a4paper,11pt]{article}
\usepackage{jheppub}
\usepackage{mathtools}

\newcommand{\dd}{\mathrm{d}}
\newcommand{\ii}{\mathrm{i}}
\newcommand{\cE}{\mathcal{E}}
\newcommand{\cK}{\mathcal{K}}
\newcommand{\cN}{\mathcal{N}}

\newcommand{\normord}[1]{:\!#1\!:}

\title{\boldmath CFT$_3$ Realizations and Celestial Representation Theory of
  the $\mathcal{L}_{\Lambda}w_{1+\infty}$ Algebra}

\hypersetup{
  pdftitle={CFT3 Realizations and Celestial Representation Theory of the Lambda-Deformed w(1+infinity) Algebra},
  pdfauthor={Bin Zhu}
}

\author[a]{Bin Zhu}
\affiliation[a]{School of Physics, Nankai University, Weijin Road 94, Tianjin 300071, P.R. China}
\emailAdd{bzhu@nankai.edu.cn}

\abstract{We study CFT$_3$ realizations and celestial representations of the
$\mathcal{L}_{\Lambda}w_{1+\infty}$ algebra.  Free scalar and Dirac theories
yield the same one-particle light-ray action.  This action extends
to the critical $O(N)$ model at $N=\infty$.  At large but finite $N$,
we use the light-ray algebra under its closure assumptions and check
prescribed iterated limits of the frequency kernel.  In
the dimension-completed celestial hard-graviton representation, a closed
raising subalgebra annihilates $G_3^+$ exactly.  We construct two nonzero
singular descendants in its induced module and prove that both vanish in
the celestial realization.  They implement the finite-spin condition and
massless spin-two shortening of the generated highest-weight sector.
The nonzero-$\Lambda$ quadratic Casimir has eigenvalue three throughout the
hard family.  The null descendants and low-spin Ward identities yield
ordered differential--difference equations for celestial correlators.}

\begin{document}
\maketitle
\flushbottom

\section{Introduction}
\label{sec:introduction}

Null-integrated local operators relate conformal field theory (CFT), collider
observables \cite{Hofman:2008ar}, and asymptotic symmetry.  The averaged null
energy (ANE) operator is the simplest example of a Lorentzian light transform
\cite{Kravchuk:2018htv}.  Its positivity follows from causality and from
modular-theoretic arguments \cite{Hartman:2016lgu,Faulkner:2016mzt}.
Under the closure assumptions of ref.~\cite{Cordova:2018ygx}, it and its
first null moment and transverse momentum flow generate a generalized
Bondi--Metzner--Sachs (BMS) algebra on a null sheet.  Products of such operators admit a light-ray operator
product expansion (OPE)
\cite{Kologlu:2019mfz}, while frequency-resolved energy flow exposes the
additional low-twist operators and regulator dependence that can enter
coincident-detector commutators \cite{Korchemsky:2021htm}.
Celestial energy-energy correlators provide a recent bridge between these
themes: ANE detectors evaluated on boost-eigenstate beam operators transform
as a celestial four-point function and are related to full-range energy
correlators by a boost-moment transform \cite{Ruan:2026xyd}.

In the celestial basis, Mellin transforms of four-dimensional scattering
states transform as two-dimensional conformal correlators
\cite{Pasterski:2017kqt}; see refs.~\cite{Donnay:2023mrd,Zhu:2026ofh} for
reviews from asymptotic-symmetry and bottom-up perspectives.
Conformally soft residues of celestial OPEs generate towers of currents
\cite{Pate:2019lpp}.  Their relation to extended BMS and celestial current
algebras was developed in refs.~\cite{Fotopoulos:2019vac,Guevara:2021abz},
and the graviton sector organizes into the wedge algebra of $w_{1+\infty}$
\cite{Strominger:2021mtt}.

The $\Lambda$-deformed algebra was introduced through a $\Lambda$-deformed
graviton OPE consistent with the Jacobi identity
\cite{Taylor:2023ajb}.  To all orders in $\Lambda$, the deformation was
subsequently identified with the
Poisson algebra of Hamiltonians on twistor space, and its Noether charges were
derived in classical self-dual gravity \cite{Bittleston:2024rqe}.  At
$\Lambda=-1$, the ANEC operator in any CFT$_3$, together with its conformal
descendants and their commutators, was shown to generate a wedge of the same
algebra \cite{Strominger:2026cft}.  More recently, $\Lambda$-corrected
asymptotic higher-spin charges were constructed on a restricted (A)dS$_4$
phase space and their algebra was verified to first order in $\Lambda$, using
charges through quadratic order in the phase-space data.  The construction
also reproduces the cosmological graviton OPE for curved conformally soft
gravitons at that order \cite{DiGiacomo:2026oku}.  These complementary
realizations motivate the explicit CFT$_3$ and celestial representation
analysis developed here.

In the CFT$_3$ construction, the C\'ordova--Shao operators form three seed
families.  We compute their canonical kernels and state the operator-domain
assumptions used in the interacting discussion.
Related CFT$_3$ constructions realize the gauge-theory $S$-algebra from
light transforms of conserved currents \cite{Sheta:2025oep}, while
stress-tensor and current light-ray operators in CFT$_4$ generate parallel
infinite-dimensional algebras \cite{Himwich:2026exq}.

We first construct explicit realizations of the three CFT$_3$ light-ray
generators in the free conformal scalar and Dirac theories.  Despite their
different field content, the two theories have the same one-particle action
and realize the same three-family sector of
$\mathcal L_{-1}w_{1+\infty}$.  We then extend the construction to the critical
$O(N)$ model, using its standard large-$N$ description and higher-spin
dictionary \cite{Moshe:2003xn,Henriksson:2022rnm,Giombi:2009wh}.  At
$N=\infty$, this gives a flavor-summed representation on the fundamental-field
sector and its singlets.

At large but finite $N$, we take the light-ray algebra as an input under
the closure assumptions stated below.  We check its compatibility with
prescribed iterated zero-frequency limits of the detector-limit kernel.
The pairwise coefficients then remain fixed while matrix elements acquire
the expected $1/N$ corrections through the dimensions of the external states.

We also analyze the hard-graviton representation of the deformed algebra at
nonzero $\Lambda$ \cite{Taylor:2023ajb}.  Related work on celestial
representations and null states includes
refs.~\cite{Banerjee:2020zlg,Banerjee:2023rni,Freidel:2022skz}.
The hard-primary action selects an exact curved highest-weight polarization
of $G_3^+$.  Its freely induced low-spin module contains two singular
descendants: a cubic polynomial in the lowering generators and the fifth
power of $P_{+,-}$.  We show that both vanish in the unprojected celestial
realization.  They implement
the finite spin-two ladder and the energy-level-one spin-one shortening
at $(E_0^{\rm AdS},s)=(3,2)$.  The dimension-completed hard family has quadratic
Casimir eigenvalue three.  Transporting the exact nulls and applying the
low-spin Ward identities gives ordered differential--difference equations,
alongside those from the highest-weight annihilators.  At $\Lambda=-1$,
the CFT$_3$ light-ray and celestial constructions realize the same
low-spin wedge sector in opposite polarizations.

This paper is organized as follows.  Section~\ref{sec:universal-algebra}
introduces the common algebra,
Section~\ref{sec:free-fields} derives its free-field realizations, and
Section~\ref{sec:critical-on} treats the critical $O(N)$ model.
Section~\ref{sec:lambda-null-state} develops the celestial representation
analysis and derives the null-decoupling equations.  The appendices collect
the canonical reductions, regulator and smearing prescriptions, projected
mode classification, and low-spin representation checks.

\section{Light-ray algebra}
\label{sec:universal-algebra}

We work in three-dimensional Minkowski space with mostly-minus signature and
null coordinates
\begin{equation}
\eta_{\mu\nu}=\operatorname{diag}(+,-,-),
\qquad
x^\pm=t\pm x,
\qquad
\dd s^2=\dd x^+\dd x^- -\dd y^2,
\qquad
g_{+-}=g_{-+}=\frac12.
\label{eq:global-metric}
\end{equation}
The null sheet is $x^-=0$, and $u\equiv x^+$ is the affine coordinate along
each complete null generator.  This is the opposite null sheet from the
mostly-plus convention of ref.~\cite{Cordova:2018ygx}.  Explicitly, with the
subscript ${\rm CS}$ denoting the C\'ordova--Shao conventions, the map
$(x^-_{\rm CS},T^{\rm CS}_{--},T^{\rm CS}_{-y})\mapsto
(u,T_{++},T_{+y})$ identifies its three light-ray operators with ours without
an additional sign.

For any conserved stress tensor, define the three light-ray densities
\begin{equation}
\begin{aligned}
\cE(y)&=\int_{-\infty}^{+\infty}\dd u\,T_{++}(u,0,y),
\\
\cK(y)&=\int_{-\infty}^{+\infty}\dd u\,uT_{++}(u,0,y),
\\
\cN(y)&=\int_{-\infty}^{+\infty}\dd u\,T_{+y}(u,0,y).
\end{aligned}
\label{eq:universal-light-ray-densities}
\end{equation}
We use the light-ray commutators of ref.~\cite{Cordova:2018ygx} under
its microcausality, unitarity, Poincare Ward-identity, and closure assumptions.
The last is a minimality hypothesis: the commutator of parallel integrated
conserved currents can again be written as an integral of a local operator.
The free-field constructions in
Section~\ref{sec:free-fields} verify the resulting algebra directly on their
common wave-packet domain.

\begin{samepage}
Applying the convention map above and specializing to one transverse
dimension gives
\begin{align}
[\cE(y_1),\cE(y_2)]&=0,
\qquad [\cK(y_1),\cK(y_2)]=0,
\nonumber\\
[\cK(y_1),\cE(y_2)]
&=-\ii\delta_{12}\cE(y_2),
\nonumber\\
[\cN(y_1),\cE(y_2)]
&=-\ii\delta_{12}\partial_{y_2}\cE(y_2)
+\ii(\partial_{y_1}\delta_{12})\cE(y_2),
\nonumber\\
[\cN(y_1),\cK(y_2)]
&=-\ii\delta_{12}\partial_{y_2}\cK(y_2)
+\ii(\partial_{y_1}\delta_{12})\cK(y_2),
\nonumber\\
[\cN(y_1),\cN(y_2)]
&=-\ii\delta_{12}\partial_{y_2}\cN(y_2)
+2\ii(\partial_{y_1}\delta_{12})\cN(y_2).
\label{eq:translated-local-signs}
\end{align}
\end{samepage}
Here $y_{12}=y_1-y_2$ and $\delta_{12}=\delta(y_{12})$.
The calligraphic symbols in
eq.~\eqref{eq:universal-light-ray-densities} are operator-valued densities on
the transverse line.  Plain symbols denote their smeared charges:
\begin{equation}
\begin{aligned}
E[f]&=\int_{\mathbb R}\dd y\,f(y)\cE(y),
& f&\in C_c^\infty(\mathbb R),
\\
K[g]&=\int_{\mathbb R}\dd y\,g(y)\cK(y),
& g&\in C_c^\infty(\mathbb R),
\\
N[Y]&=\int_{\mathbb R}\dd y\,Y(y)\cN(y),
& Y&\in C_c^\infty(\mathbb R).
\end{aligned}
\label{eq:universal-smeared-definitions}
\end{equation}
Here $Y$ denotes the $y$ component of a transverse vector field.  Smearing the
full set of local commutators in eq.~\eqref{eq:translated-local-signs} and
integrating the transverse delta-function derivatives by parts gives
\begin{align}
[E[f],E[g]]&=0,
&
[K[f],K[g]]&=0,
&
[K[g],E[f]]&=-\ii E[gf],
\nonumber\\
[N[Y],E[f]]&=\ii E[Yf'],
&
[N[Y],K[g]]&=\ii K[Yg'],
&
[N[Y],N[Z]]&=\ii N[YZ'-ZY'].
\label{eq:universal-smeared-algebra}
\end{align}
Here a prime denotes differentiation with respect to $y$.
In addition to the closure assumptions just stated,
eq.~\eqref{eq:universal-smeared-algebra} uses the null endpoint and smearing
conditions in Appendix~\ref{app:regulators}.

To obtain a compact mode basis directly from the null-plane algebra, write
$\cE_y\equiv\cE$, $\cK_y\equiv\cK$, and $\cN_y\equiv\cN$ for the line
densities and use $y=\tan(\varphi/2)$ as a coordinate compactification of
the transverse line.  Then $\varphi\sim\varphi+2\pi$ and
$J(\varphi)=\dd y/\dd\varphi$.  With the subscript ${\rm circ}$ denoting this
algebraic circle basis, we define
\begin{equation}
\cE_{\rm circ}=J\cE_y,
\qquad
\cK_{\rm circ}=J\cK_y,
\qquad
\cN_{\rm circ}=J^2\cN_y,
\qquad
Y^\varphi=J^{-1}Y^y,
\label{eq:line-cylinder-densities}
\end{equation}
so that
$E[f]=\int\dd\varphi\,f\cE_{\rm circ}$,
$K[g]=\int\dd\varphi\,g\cK_{\rm circ}$, and
$N[Y]=\int\dd\varphi\,Y^\varphi\cN_{\rm circ}$ are equal to their line
expressions.

These definitions implement the coordinate push-forward of the null-plane
charges.  The conformal map from Minkowski space to the Einstein cylinder
in ref.~\cite{Strominger:2026cft} additionally assigns conformal weights and
mixes the transverse momentum flow with the first null moment.
The Fourier modes are the special
smearings
\begin{equation}
E_k=\int_0^{2\pi}\dd\varphi\,e^{\ii k\varphi}\cE_{\rm circ},
\qquad
K_k=\int_0^{2\pi}\dd\varphi\,e^{\ii k\varphi}\cK_{\rm circ},
\qquad
N_k=\int_0^{2\pi}\dd\varphi\,e^{\ii k\varphi}\cN_{\rm circ},
\quad k\in\mathbb Z.
\label{eq:universal-cylinder-modes}
\end{equation}
Writing $e_k(y)=e^{\ii k\varphi(y)}$, the same modes expressed as line
smearings are
\begin{equation}
E_k=E[e_k],
\qquad
K_k=K[e_k],
\qquad
N_k=N[Je_k],
\qquad
\partial_y=J^{-1}\partial_\varphi.
\label{eq:universal-mode-pullback}
\end{equation}
Using eq.~\eqref{eq:universal-smeared-algebra} with the pulled-back modes in
eq.~\eqref{eq:universal-mode-pullback} gives
\begin{align}
[E_k,E_l]&=0,
&
[K_k,K_l]&=0,
&
[K_k,E_l]&=-\ii E_{k+l},
\nonumber\\
[N_k,E_l]&=-lE_{k+l},
&
[N_k,K_l]&=-lK_{k+l},
&
[N_k,N_l]&=(k-l)N_{k+l}.
\label{eq:universal-mode-algebra}
\end{align}
Reference~\cite{Strominger:2026cft} obtains the same abstract mode algebra for
a geometrically distinguished set
of light rays on the Einstein cylinder.  Its conformal map supplies the
$SO(3,2)$ interpretation of those modes, whereas the construction above is an
algebraic realization on the compactified transverse line.  We use the
isomorphism between these two mode algebras below.

The $\mathcal{L}_{\Lambda}w_{1+\infty}$ algebra is defined by the commutation
relations \cite{Taylor:2023ajb,Strominger:2026cft}
\begin{align}
[w^p_{\bar m,m},w^q_{\bar n,n}]
&=\big[\bar m(q-1)-\bar n(p-1)\big]\,
w^{p+q-2}_{\bar m+\bar n,m+n}
\nonumber\\
&\quad
-\Lambda\big[m(q-2)-n(p-2)\big]\,
w^{p+q-1}_{\bar m+\bar n,m+n}.
\label{eq:Lambda-w-bracket}
\end{align}
Here $p,q$ label generator families, while $(\bar m,m)$ and
$(\bar n,n)$ are their two mode-index pairs; the bars on $\bar m$ and
$\bar n$ are labels.  The three-family light-ray sector and the celestial
sector use the index lattices fixed below by eqs.~\eqref{eq:universal-w-map}
and~\eqref{eq:null-soft-w-map}, respectively.
Through the mode-algebra isomorphism just described, we adopt the
$w$-generator labels of ref.~\cite{Strominger:2026cft}.  For the three
stress-tensor families, define
\begin{align}
W_E(k)&=w^{\frac{k+3}{2}}_{-\frac{k+1}{2},\,\frac{k-1}{2}},
&
E_k&=-\ii^kW_E(k),
\nonumber\\
W_+(k)&=w^{1+\frac{k}{2}}_{-\frac{k}{2},\,\frac{k}{2}},
&
N_k+\ii K_k&=-2\ii^kW_+(k),
\nonumber\\
W_-(k)&=w^{2+\frac{k}{2}}_{-\frac{k}{2},\,\frac{k}{2}},
&
N_k-\ii K_k&=-2\ii^kW_-(k).
\label{eq:universal-w-map}
\end{align}
Substituting the identifications in eq.~\eqref{eq:universal-w-map} into
eq.~\eqref{eq:Lambda-w-bracket} at $\Lambda=-1$ reproduces
eq.~\eqref{eq:universal-mode-algebra}.
Equation~\eqref{eq:universal-w-map} identifies the three C\'ordova--Shao seed
families.  In the
Einstein-cylinder construction of ref.~\cite{Strominger:2026cft}, the ANEC
modes, their descendants under the global $SO(3,2)$ charges, and commutators
of those descendants generate one operator at every lattice point satisfying
\begin{equation}
\bar m+p\geq1,
\qquad
m-p\geq-2,
\qquad
p\pm m,\ p\pm\bar m\in\mathbb Z.
\label{eq:universal-cft3-wedge}
\end{equation}
The full conformal action thus completes the seed sector to the CFT$_3$ wedge.
The generic-$\Lambda$ bracket will be used again in
Section~\ref{sec:lambda-null-state}.

\section{Free-field realizations}
\label{sec:free-fields}

\subsection{Conformal scalar}
\label{sec:free-scalar}

For the free scalar, the stress-tensor light-ray generators can be evaluated
directly on Fock space.  Their normal-ordered oscillator kernels realize the
three-family sector of $\mathcal{L}_{\Lambda}w_{1+\infty}$ at the physical
value $\Lambda=-1$.

Consider a real conformal scalar with Lorentzian action and Euclidean
normalization
\begin{equation}
S_\phi=\frac12\int\dd^3x\,\partial_\mu\phi\,\partial^\mu\phi,
\qquad
\Box\phi=0,
\qquad
\langle\phi(x)\phi(0)\rangle_{\mathrm E}=\frac{1}{4\pi|x|}.
\label{eq:scalar-action}
\end{equation}
The conformally improved stress tensor is
\begin{equation}
T_{\mu\nu}=\partial_\mu\phi\,\partial_\nu\phi
-\frac12\eta_{\mu\nu}(\partial\phi)^2
+\frac18\left(\eta_{\mu\nu}\Box
-\partial_\mu\partial_\nu\right)\phi^2.
\label{eq:scalar-stress}
\end{equation}
It obeys $\partial^\mu T_{\mu\nu}=(\partial_\nu\phi)\Box\phi$ and
$T^\mu{}_{\mu}=\phi\Box\phi/2$, and is therefore conserved and traceless at
separated points on shell.  Since $g_{++}=g_{+y}=0$ and
$\partial_+=\partial_u$ on the null sheet, selecting the $++$ and $+y$
components of eq.~\eqref{eq:scalar-stress} gives
\begin{equation}
T_{++}=(\partial_u\phi)^2-\frac18\partial_u^2\phi^2,
\qquad
T_{+y}=\partial_u\phi\,\partial_y\phi
-\frac18\partial_u\partial_y\phi^2.
\label{eq:scalar-null-stress}
\end{equation}
This component reduction holds off shell.  The mixed
component is required because its complete-null-line integral generates
transverse reparametrizations in the algebra below.

The scalar charges are the universal operators in
eq.~\eqref{eq:universal-light-ray-densities}.  After the null-sheet exchange
below eq.~\eqref{eq:global-metric}, they are the
operators $\mathcal E$, $\mathcal K$, and $\mathcal N_y$ of
ref.~\cite{Cordova:2018ygx}.  We define them on wave-packet states for which
$\phi^2$, $\partial_u\phi^2$, and $u\partial_u\phi^2$ vanish at null infinity.
The improvement terms then integrate to zero.  The endpoint identities and
the distributional origin of the $1/2$ term in the first null moment are
derived in Appendix~\ref{app:canonical-details}.

Let $a(p,y)$, with longitudinal momentum $p>0$, denote the scalar annihilation
operator in the mixed longitudinal-momentum/transverse-position basis used in
Appendix~\ref{app:canonical-details}.  To realize the algebraic circle basis
of Section~\ref{sec:universal-algebra}, we use $y=\tan(\varphi/2)$ and
normalize its transverse half-density with respect to $\dd\varphi/(2\pi)$,
\begin{equation}
\begin{aligned}
a(p,\varphi)&=\left(2\pi\frac{\dd y}{\dd\varphi}\right)^{1/2}a(p,y),
\\
[a(p,\varphi),a^\dagger(p',\varphi')]
&=2\pi\delta(p-p')\delta_{2\pi}(\varphi-\varphi'),
\\
\dd\mu&=\frac{\dd p}{2\pi}\frac{\dd\varphi}{2\pi},
\end{aligned}
\label{eq:scalar-cylinder-map}
\end{equation}
where the periodic delta function is normalized by
$\int_0^{2\pi}\dd\varphi\,\delta_{2\pi}(\varphi-\varphi')
f(\varphi)/(2\pi)=f(\varphi')$.  With $k\in\mathbb Z$, the circle modes are
\begin{align}
E_k&=\int\dd\mu\,
\normord{a^\dagger e^{\ii k\varphi}p\,a},
\nonumber\\
K_k&=\int\dd\mu\,
\normord{a^\dagger e^{\ii k\varphi}
\left[-\ii\left(p\partial_p+\frac12\right)\right]a},
\nonumber\\
N_k&=\int\dd\mu\,
\normord{a^\dagger
\left[\ii e^{\ii k\varphi}\partial_\varphi
-\frac{k}{2}e^{\ii k\varphi}\right]a}.
\label{eq:scalar-EKN-modes}
\end{align}
Derivatives act on the annihilation operator to their right.  The term
$-k/2$ is the connection of a transverse half-density and follows by
integrating the symmetric momentum density by parts.  It also makes
$E_k^\dagger=E_{-k}$, $K_k^\dagger=K_{-k}$, and $N_k^\dagger=N_{-k}$.

The charges in eq.~\eqref{eq:scalar-EKN-modes} are second quantizations of
one-particle operators.  For
\begin{equation}
Q[A]=\int\dd\mu\,\normord{a^\dagger A a},
\qquad
[Q[A],Q[B]]=Q([A,B]),
\label{eq:scalar-bilinear-lift}
\end{equation}
the second identity follows directly from the canonical oscillator
commutator.  On the wave-packet domain used above, with the $p=0$ mode
excluded until the regulator is removed, these number-conserving bilinears
produce no central term.

Writing $z=e^{\ii\varphi}$, their one-particle kernels are
\begin{equation}
\widehat E_k=z^kp,
\qquad
\widehat K_k=-\ii z^k\left(p\partial_p+\frac12\right),
\qquad
\widehat N_k=-z^{k+1}\partial_z-\frac{k}{2}z^k.
\label{eq:scalar-differential-kernels}
\end{equation}
For instance, acting on a one-particle test wavefunction $f(p,z)$,
\begin{align}
[\widehat K_k,\widehat E_l]f
&=-\ii z^{k+l}
\left[\left(p\partial_p+\frac12\right)(pf)
-p\left(p\partial_p+\frac12\right)f\right]
\nonumber\\
&=-\ii\widehat E_{k+l}f.
\label{eq:scalar-KE-derivation}
\end{align}
Together with the other five kernel commutators, this reproduces the canonical
six relations in eq.~\eqref{eq:universal-mode-algebra}.  The Jacobi identities
follow from the associative one-particle operator algebra and
eq.~\eqref{eq:scalar-bilinear-lift}.  Representative Jacobi and contact-term
checks are given in Appendix~\ref{app:canonical-details}.

The embedding \eqref{eq:universal-w-map} then implies
\begin{align}
[W_E(k),W_+(l)]
&=\frac{\ii^{-(k+l)}}{2}[E_k,N_l+\ii K_l]
\nonumber\\
&=\frac{1-k}{2}W_E(k+l),
\label{eq:scalar-w-representative-check}
\end{align}
which agrees with eq.~\eqref{eq:Lambda-w-bracket} at $\Lambda=-1$.  The
other five independent pairings are listed in
Appendix~\ref{app:canonical-details} and reproduce the abstract bracket.

Thus, on the normal-ordered wave-packet domain, the oscillator bilinears
realize the light-ray algebra \eqref{eq:universal-mode-algebra} and its
three-family $\Lambda=-1$ embedding.  This agrees with the free-scalar
correlation-function analysis of
ref.~\cite{Cordova:2018ygx} and the frequency-resolved free-field light-ray
algebra of ref.~\cite{Korchemsky:2021htm}.  The canonical reduction provides
the normalization benchmark for the fermion and critical $O(N)$ theories.

\subsection{Dirac fermion}
\label{sec:free-fermion}

The fermion theory separates the role of statistics from that of the scalar
improvement term.  Solving the light-front constraint reduces the
stress-tensor moments to particle and antiparticle bilinears, whose
commutators follow from the canonical anticommutation relations (CAR).

We use a complex two-component Dirac fermion with
\begin{equation}
\gamma^0=\sigma^3,
\qquad
\gamma^1=\ii\sigma^1,
\qquad
\gamma^2=\ii\sigma^2,
\qquad
\{\gamma^\mu,\gamma^\nu\}=2\eta^{\mu\nu},
\label{eq:fermion-gamma-conventions}
\end{equation}
where $\sigma^i$ are the Pauli matrices, and
$\bar\psi=\psi^\dagger\gamma^0$.  The explicitly Hermitian action is
\begin{equation}
S_\psi=\frac{\ii}{2}\int\dd^3x\,
\bar\psi\gamma^\mu\overleftrightarrow{\partial_\mu}\psi,
\qquad
\gamma^\mu\partial_\mu\psi=0,
\qquad
(\partial_\mu\bar\psi)\gamma^\mu=0.
\label{eq:fermion-action-eom}
\end{equation}
Here
$\bar\psi\gamma^\mu\overleftrightarrow{\partial_\mu}\psi
=\bar\psi\gamma^\mu\partial_\mu\psi
-(\partial_\mu\bar\psi)\gamma^\mu\psi$.

The symmetric stress tensor is
\begin{align}
T_{\mu\nu}^{\psi}
&=\frac{\ii}{4}\bar\psi
\left(\gamma_\mu\overleftrightarrow{\partial_\nu}
+\gamma_\nu\overleftrightarrow{\partial_\mu}\right)\psi.
\label{eq:fermion-stress}
\end{align}
In the coordinates of eq.~\eqref{eq:global-metric},
$\gamma_+=(\gamma^0-\gamma^1)/2$, $\gamma_y=-\gamma^2$, and
$\partial_+=\partial_u$.  Selecting the two components needed in
eq.~\eqref{eq:universal-light-ray-densities} therefore gives
\begin{equation}
T_{++}^{\psi}
=\frac{\ii}{2}\bar\psi\gamma_+
\overleftrightarrow{\partial_u}\psi,
\qquad
T_{+y}^{\psi}
=\frac{\ii}{4}\bar\psi
\left(\gamma_+\overleftrightarrow{\partial_y}
+\gamma_y\overleftrightarrow{\partial_u}\right)\psi.
\label{eq:fermion-stress-null-components}
\end{equation}
Both terms in the mixed component are required: on the light front they
contribute equally to the transverse momentum-flow kernel.  The equations of
motion imply $T^\mu{}_{\mu}=0$ and $\partial^\mu T_{\mu\nu}=0$.  The
light-front reduction is given in Appendix~\ref{app:canonical-details}.  Let
$c(p,y)$ and $d(p,y)$ denote the corresponding particle and antiparticle
annihilation operators and write $s\in\{c,d\}$.  Suppressing their common
$(p,y)$ arguments
inside normal-ordered products, the energy density and the remaining two
one-particle kernels reduce to
\begin{equation}
\cE_\psi(y)
=\sum_{s=c,d}\int_0^\infty\frac{\dd p}{2\pi}\,
p\normord{s^\dagger s},
\qquad
\widehat K_\psi=-\ii\left(p\partial_p+\frac12\right),
\qquad
\widehat N_\psi[Y]=\ii\left(Y\partial_y+\frac12Y'\right).
\label{eq:fermion-reduced-kernels}
\end{equation}
The first expression is positive, while the other two reproduce the scalar
contact term and half-density connection.  After the line-to-circle map,
$N_0^\psi$ generates translations of the circle coordinate $\varphi$, and
$(N_k^\psi)^\dagger=N_{-k}^\psi$.  It is the line smearing $N^\psi[J]$,
not the uniform line translation $N^\psi[1]$.  Since
$J=(1+y^2)/2$ and $J^{-1}=1+\cos\varphi$,
\begin{equation}
N^\psi[1]=N_0^\psi+\frac12\left(N_1^\psi+N_{-1}^\psi\right).
\label{eq:fermion-line-translation}
\end{equation}
Using the oscillator normalization and
measure in eq.~\eqref{eq:scalar-cylinder-map} for both $c$ and $d$, the modes are
\begin{align}
E_k^\psi
&=\sum_{s=c,d}\int\dd\mu\,
\normord{s^\dagger e^{\ii k\varphi}p\,s},
\nonumber\\
K_k^\psi
&=\sum_{s=c,d}\int\dd\mu\,
\normord{s^\dagger e^{\ii k\varphi}
\left[-\ii\left(p\partial_p+\frac12\right)\right]s},
\nonumber\\
N_k^\psi
&=\sum_{s=c,d}\int\dd\mu\,
\normord{s^\dagger
\left[\ii e^{\ii k\varphi}\partial_\varphi
-\frac{k}{2}e^{\ii k\varphi}\right]s}.
\label{eq:fermion-EKN-modes}
\end{align}
Their one-particle operators are therefore the half-density kernels in
eq.~\eqref{eq:scalar-differential-kernels}.

For an even fermion bilinear
\begin{equation}
Q_\psi[A]=\sum_{s=c,d}\int\dd\mu\,
\normord{s^\dagger A s},
\qquad
[Q_\psi[A],Q_\psi[B]]=Q_\psi([A,B]).
\label{eq:fermion-bilinear-lift}
\end{equation}
the second identity follows from the CAR.  Applying
eq.~\eqref{eq:fermion-bilinear-lift} to
eq.~\eqref{eq:fermion-EKN-modes} reproduces all six canonical relations in
eq.~\eqref{eq:universal-mode-algebra}.  No statistics-dependent sign remains
in the commutator of the even charges.
The contact-term check in Appendix~\ref{app:canonical-details} shows that no
additional central term is generated on the wave-packet domain specified in
Section~\ref{sec:free-scalar}.

Because the fermion and scalar kernels coincide, the common map
\eqref{eq:universal-w-map} gives the same three-family brackets at
$\Lambda=-1$, while eq.~\eqref{eq:fermion-bilinear-lift} implies the Jacobi
identities.  The two local reductions yield the same one-particle
representation.  This agrees with the frequency-resolved free-fermion
light-ray algebra of ref.~\cite{Korchemsky:2021htm}.

\section{Critical \texorpdfstring{$O(N)$}{O(N)} model}
\label{sec:critical-on}

The critical $O(N)$ model distinguishes an explicit leading Fock-space
representation from an interacting application of the light-ray algebra.
We construct the leading representation using the fundamental fields.  At finite $N$,
we assume the closure conditions of Section~\ref{sec:universal-algebra},
check iterated limits of the frequency kernel, and evaluate
Ward-algebra matrix elements.

\subsection{Leading large-\texorpdfstring{$N$}{N} representation}
\label{sec:on-large-N-setup}

For Euclidean CFT data we use the Hubbard--Stratonovich description
\cite{Giombi:2009wh,Henriksson:2022rnm}
\begin{equation}
S_{\mathrm{HS}}=\int\dd^3x\left[
\frac12\partial_\mu\phi^i\partial^\mu\phi^i
+\frac{1}{2\sqrt N}\sigma\phi^i\phi^i\right],
\qquad i=1,\ldots,N.
\label{eq:on-HS-action}
\end{equation}
Here $\phi^i$ is the fundamental $O(N)$ vector and $\sigma$ is the singlet
Hubbard--Stratonovich field.
We work at the infrared critical point, with the mass deformation tuned to
criticality and the irrelevant local $\sigma^2$ term omitted.
The Lorentzian light-ray operators are obtained by analytic continuation to
the signature and null coordinates of eq.~\eqref{eq:global-metric}.  With the standard
large-$N$ contour, the fundamental-field bubble fixes the collective
normalization to
\begin{equation}
\langle\sigma(x)\sigma(0)\rangle
=\frac{16}{\pi^2|x|^4}+O(N^{-1}),
\qquad
g_{\sigma\phi\phi}=O(N^{-1/2}).
\label{eq:on-sigma-normalization}
\end{equation}
Here $g_{\sigma\phi\phi}$ denotes the $\sigma\phi\phi$ vertex, equivalently
the associated three-point coupling; only its large-$N$ scaling is used
below.  The dimension and stress-tensor data in the same convention are
\cite{Henriksson:2022rnm,Petkou:1994ad}
\begin{align}
\Delta_\phi
&=\frac12+\frac{4}{3\pi^2N}+O(N^{-2}),
&
\Delta_\sigma
&=2-\frac{32}{3\pi^2N}+O(N^{-2}),
\nonumber\\
C_T
&=\frac{3N}{32\pi^2}
\left(1-\frac{40}{9\pi^2N}+O(N^{-2})\right).
\label{eq:on-large-N-data}
\end{align}
Here $C_T$ is defined by
$\langle T_{\mu\nu}(x)T_{\rho\sigma}(0)\rangle
=C_T\mathcal I_{\mu\nu,\rho\sigma}(x)/|x|^6$, where $\mathcal I$ is the
standard dimensionless conformal tensor structure; at leading order $C_T$ is $N$
times the real-scalar value.

We distinguish two singlet spaces.  The leading fundamental-field space
$\mathcal H_{\phi,\mathrm{sing}}^{(0)}$ consists of finite-particle
wave packets built from $a_i^\dagger$ with all flavor indices contracted into
$O(N)$ invariants; descendants and arbitrary finite multi-particle singlets
are included.  The oscillator construction below acts on
$\mathcal H_{\phi,\mathrm{sing}}^{(0)}$.  The full critical singlet sector
additionally contains the collective primary $\sigma$, its descendants,
and their multi-traces.

The leading stress tensor is correspondingly
\begin{equation}
T_{\mu\nu}^{(0)}=\sum_{i=1}^N\left[
\partial_\mu\phi^i\partial_\nu\phi^i
-\frac12\eta_{\mu\nu}(\partial\phi^i)^2
+\frac18(\eta_{\mu\nu}\Box-\partial_\mu\partial_\nu)(\phi^i)^2
\right].
\label{eq:on-leading-stress}
\end{equation}
At finite $N$, $T_{\mu\nu}$ denotes the unique renormalized conserved
spin-two primary fixed by translation Ward identities.  A local separation
into orders in $1/N$ depends on the renormalization scheme and on whether
$\sigma$ is retained as an explicit field.

Let $a_i(p,\varphi)$ obey the scalar commutator of
eq.~\eqref{eq:scalar-cylinder-map}, with an additional $\delta_{ij}$.  In
terms of the scalar kernels in eq.~\eqref{eq:scalar-differential-kernels}, the
leading representation is
\begin{align}
Q_N[A]&=\sum_{i=1}^N\int\dd\mu\,
\normord{a_i^\dagger A a_i},
\nonumber\\
(E_k^{(0)},K_k^{(0)},N_k^{(0)})
&=(Q_N[\widehat E_k],Q_N[\widehat K_k],Q_N[\widehat N_k]),
\nonumber\\
[Q_N[A],Q_N[B]]&=Q_N([A,B]),
\label{eq:on-leading-representation}
\end{align}
where the species Kronecker delta removes all cross terms.  The scalar
endpoint conditions apply component by component, and the common
half-density kernels therefore imply the canonical algebra
\eqref{eq:universal-mode-algebra} with
$(E,K,N)\mapsto(E^{(0)},K^{(0)},N^{(0)})$.  No power of $N$ appears in the
structure constants, despite every generator being a sum over $N$ fields.
Thus the algebra closes on the finite-particle wave-packet Fock domain and on
its invariant subspace $\mathcal H_{\phi,\mathrm{sing}}^{(0)}$.

\subsection{Finite-\texorpdfstring{$N$}{N} consistency checks}
\label{sec:on-interaction-corrections}

At finite $N$ we use the algebra
\eqref{eq:universal-smeared-algebra} under its stated closure and domain
assumptions and test its compatibility with prescribed iterated limits of
the detector kernel.  Formally, the moment map is
\begin{equation}
\cE_\omega(y)=\int_{-\infty}^{+\infty}\dd u\,
e^{-\ii\omega u}T_{++}(u,0,y),
\qquad
\cE=\cE_\omega\big|_{\omega=0},
\qquad
\cK=\ii\partial_\omega\cE_\omega\big|_{\omega=0}.
\label{eq:on-frequency-moment-map}
\end{equation}
With a smooth finite null-time cutoff, these operations define regulated
moments.  Their removal requires control of matrix elements, as discussed
in Appendix~\ref{app:regulators}.  Separating the two null sheets by a
distance $\epsilon$, let
$\widetilde n_\phi=n_\phi C_T^{(\phi)}/C_T$ be the scalar weight in the
parity-even three-stress-tensor ($TTT$) structure.  Here $n_\phi$ is the
coefficient of the free-scalar tensor structure, $C_T^{(\phi)}$ is the
real-scalar two-point normalization, and $f_\phi$ denotes the corresponding
light-transform kernel given in
eq.~\eqref{eq:on-scalar-frequency-structure}.  The coincident-detector limit
of the generalized energy-flow commutator has the form
\cite{Korchemsky:2021htm}
\begin{align}
&[\cE_{\omega_1}(y_1),\cE_{\omega_2}(y_2)]_\epsilon
\nonumber\\
&\xrightarrow[\epsilon\to0^+]{}\delta(y_{12})\Big[
\big(\omega_2-\omega_1
+\widetilde n_\phi\omega_1^2\omega_2^2
f_\phi(\omega_1,\omega_2)\big)
\cE_{\omega_1+\omega_2}(y_2)
\nonumber\\
&\hspace{19mm}
{}-C\omega_1^3\delta(\omega_1+\omega_2)\mathbf1
+\sum_{\mathcal O:\,1/2\le\Delta_{\mathcal O}\le1}
c_{TT\mathcal O}f_{\mathcal O}(\omega_1,\omega_2)
\mathcal O_{\omega_1+\omega_2}(y_2)\Big].
\label{eq:on-frequency-commutator}
\end{align}
Here $\mathbf1$ is the identity operator, $C$ is its regulator-dependent
coefficient,
$\mathcal O_\omega$ is the frequency transform of a scalar primary,
$c_{TT\mathcal O}$ is its $TT$ OPE coefficient in the chosen two-point
normalization, and $f_{\mathcal O}$ is the corresponding light-transform
kernel.  The displayed terms are precisely those that survive an isolated
pairwise detector limit.  Operators of twist greater than one are suppressed
in this limit but can re-enter nested regulated commutators.  In $d=3$ the
parity-odd $\langle TTT\rangle$ structure does not contribute to
eq.~\eqref{eq:on-frequency-commutator} \cite{Korchemsky:2021htm}.

In three dimensions, the leading critical-model value is
$\widetilde n_\phi=1+O(N^{-1})$.  For the kernel check, put
$\omega_1=x>0$, $\omega_2=-s<0$ and differentiate at strictly positive
$x,s$.  We then take $s\to0^+$ at fixed $x>0$, followed by $x\to0^+$.
The explicit kernel in Appendix~\ref{app:regulators} obeys
\begin{equation}
\lim_{x\to0^+}\lim_{s\to0^+}
\left\{1,\partial_x,\partial_x\partial_s\right\}
\big[x^2s^2f_\phi(x,-s)\big]=\{0,0,0\}.
\label{eq:on-scalar-frequency-moments}
\end{equation}
These are the scalar-kernel contributions to the prescribed $EE$, $KE$,
and $KK$ operations, up to the Fourier factor in $KE$.
The order-$1/N$ correction to $\widetilde n_\phi$ does not affect these
zeros.  Products with a matrix element obey the same axis limits if that
matrix element and its required frequency derivatives remain finite there.
Extension to the full regulated matrix elements requires uniform bounds
on these quantities as the sheet regulator is removed.

The limit is nonuniform at the frequency origin.  Writing
$F(x,s)=x^2s^2f_\phi(x,-s)$, the exact expression
\eqref{eq:on-scalar-frequency-closed} gives
$F(t,t)=t$ and $\partial_x\partial_sF(t,t)=1/(2t)$.
The derivative limits therefore depend on the approach to the origin.
Appendix~\ref{app:regulators} also checks
the displayed central distribution after smooth finite-time regularization.

The remaining scalar sum in eq.~\eqref{eq:on-frequency-commutator} is empty in
the critical singlet channel in the controlled large-$N$ regime.  The
fundamental field is not an $O(N)$ singlet, while
eq.~\eqref{eq:on-large-N-data} gives $\Delta_\sigma>1$.  In particular, the
$TT\sigma$ term is multiplied by
$\epsilon^{(\Delta_\sigma-1)/2}$ and vanishes in the isolated fixed-$N$
detector limit at fixed nonzero frequencies, irrespective of the value
of $c_{TT\sigma}$.  Bounds on the frequency derivatives before removing
$\epsilon$ require control of the $TT\sigma$ and heavier-singlet
contributions.  Broken
higher-spin currents other than the stress tensor have twist
$\tau_s=1+\gamma_s/N+O(N^{-2})$ \cite{Henriksson:2022rnm}, where $s$ is the
spin and $\gamma_s>0$ is the leading anomalous-twist coefficient.  They vanish
in the isolated pairwise detector limit if that limit is taken at fixed $N$.
Reversing the two limits gives a different expansion:
\begin{equation}
\begin{aligned}
\lim_{\epsilon\to0^+}\epsilon^{\gamma_s/(2N)}
&=0
&& (N<\infty,\ \gamma_s>0),
\\
\epsilon^{\gamma_s/(2N)}
&=1+\frac{\gamma_s}{2N}\log\epsilon
+O\!\left(\frac{(\log\epsilon)^2}{N^2}\right)
&& (N\to\infty,\ \epsilon>0\ \text{fixed}).
\end{aligned}
\label{eq:on-noncommuting-limits}
\end{equation}
The noncommutativity in eq.~\eqref{eq:on-noncommuting-limits} concerns
$\epsilon$ and $N$.  The regulator question involving frequency
differentiation is formulated in Appendix~\ref{app:regulators}.
Under the light-ray closure assumptions, the pairwise relations are
\begin{equation}
[\cE(y_1),\cE(y_2)]=0,
\qquad
[\cK(y_1),\cE(y_2)]=-\ii\delta(y_{12})\cE(y_2),
\qquad
[\cK(y_1),\cK(y_2)]=0.
\label{eq:on-local-EEK-algebra}
\end{equation}
The universal term in eq.~\eqref{eq:on-frequency-commutator} and the
prescribed iterated kernel limits are compatible with these relations.
The coefficient-level Jacobi identities for the six brackets follow from
eq.~\eqref{eq:universal-mode-algebra}.  Regulated nested commutators can
probe operators suppressed in an isolated pairwise detector limit.

With the light-ray algebra as input, interactions correct its matrix
elements through the external-state dimensions.  For a scalar primary
$\mathcal O$ of dimension $\Delta$ inserted at the Euclidean-regulated
positions $t=\pm\ii\delta$, the stress-tensor Ward identity gives
\begin{equation}
\frac{\langle\mathcal O(\ii\delta)\cE(y)
\mathcal O(-\ii\delta)\rangle}
{\langle\mathcal O(\ii\delta)\mathcal O(-\ii\delta)\rangle}
=\frac{\Delta\delta^2}{\pi(\delta^2+y^2)^2},
\qquad
\int\dd y\,\langle\cE(y)\rangle_{\mathcal O,\delta}
=\frac{\Delta}{2\delta}.
\label{eq:on-one-ray-profile}
\end{equation}
Here $\langle\cdots\rangle_{\mathcal O,\delta}$ denotes the normalized
matrix element defined by the ratio in the first part of the equation.
The functional form follows by performing the complete null integral of the
scalar--scalar--stress-tensor three-point function.  Its normalization follows
from the translation Ward identity together with
$\int_{-\infty}^{\infty}\dd y\,\delta^2/(\delta^2+y^2)^2=\pi/(2\delta)$,
which gives the second relation in eq.~\eqref{eq:on-one-ray-profile}.
In the reflection-symmetric state,
$\langle\cK(y)\rangle=\langle\cN(y)\rangle=0$.  The compactly supported
test functions $\chi_R$ and $h_R=y\chi_R$ are defined in
eq.~\eqref{eq:on-regulated-test-functions}.  Their transverse cutoff
limits converge by the $|y|^{-4}$ falloff of
eq.~\eqref{eq:on-one-ray-profile}.  Combining the algebra
\eqref{eq:universal-smeared-algebra}, the one-ray Ward identity, and the
large-$N$ data \eqref{eq:on-large-N-data} gives
\begin{align}
\lim_{R\to\infty}
\langle[K[\chi_R],E[\chi_R]]\rangle_{\phi,\delta}
&=-\ii\left(\frac{1}{4\delta}
+\frac{2}{3\pi^2N\delta}\right)+O(N^{-2}),
\nonumber\\
\lim_{R\to\infty}
\langle[N[\chi_R],E[h_R]]\rangle_{\phi,\delta}
&=+\ii\left(\frac{1}{4\delta}
+\frac{2}{3\pi^2N\delta}\right)+O(N^{-2}),
\nonumber\\
\lim_{R\to\infty}
\langle[K[\chi_R],E[\chi_R]]\rangle_{\sigma,\delta}
&=-\ii\left(\frac{1}{\delta}
-\frac{16}{3\pi^2N\delta}\right)+O(N^{-2}),
\nonumber\\
\lim_{R\to\infty}
\langle[N[\chi_R],E[h_R]]\rangle_{\sigma,\delta}
&=+\ii\left(\frac{1}{\delta}
-\frac{16}{3\pi^2N\delta}\right)+O(N^{-2}).
\label{eq:on-NLO-probes}
\end{align}
The subscripts $(\phi,\delta)$ and $(\sigma,\delta)$ denote the normalized
regulated matrix elements with external $\phi^i$ and $\sigma$ insertions,
respectively.
The $1/N$ corrections enter through the external-state dimensions, while
the pairwise algebra coefficients remain fixed under the closure
assumptions.  The $\phi^i$ and $\sigma$ matrix elements probe the
nonsinglet and singlet Ward actions, respectively.

\section{Celestial null states and differential--difference equations}
\label{sec:lambda-null-state}

The preceding sections realize a light-ray sector of the deformed algebra
in CFT$_3$ at $\Lambda=-1$.  We now study its action on positive-helicity
celestial hard gravitons at nonzero $\Lambda$.  The deformed translations
select an exact highest-weight state $G_3^+$ and link it to hard states of
other Mellin dimensions.  We construct two singular descendants of this
highest-weight state, prove that both vanish in the celestial realization,
and derive their differential--difference equations from the low-spin Ward
identities.

\subsection{Deformed hard-primary action and highest-weight conditions}
\label{sec:lambda-null-hard-action}
\label{sec:lambda-closed-polarization}

Here $\Delta$ is the Mellin dimension of a positive-helicity hard graviton,
with weights
\begin{equation}
h=\frac{\Delta+2}{2},
\qquad
\bar h=\frac{\Delta-2}{2}.
\label{eq:null-hard-weights}
\end{equation}
We use radial-quantization notation $\lvert G_\delta^+\rangle$ for an
insertion at the celestial origin.  The state space includes its
integer-shifted Mellin labels and global descendants.  These are formal
hard-state vectors, including at integer $\delta$; no additional conformally
soft residue or inner-product quotient is taken.  Our residue convention is
\begin{equation}
L_{-1}\lvert G_\delta^+\rangle=\partial\lvert G_\delta^+\rangle,
\qquad
\bar L_{-1}\lvert G_\delta^+\rangle=-\bar\partial\lvert G_\delta^+\rangle.
\label{eq:null-derivative-convention}
\end{equation}

The low-spin identification in the index ordering of
eq.~\eqref{eq:Lambda-w-bracket} is \cite{Taylor:2023ajb}
\begin{equation}
P_{a,b}=w^{3/2}_{b,a},
\qquad a,b\in\left\{-\frac12,\frac12\right\},
\qquad
w^1_{0,m}=\Lambda L_m,
\qquad
w^2_{n,0}=\bar L_n,
\qquad m,n\in\{-1,0,1\}.
\label{eq:null-translation-identification}
\end{equation}
A sign in $P_{\pm,\pm}$ denotes $\pm1/2$.  We denote the ten-generator
algebra by $\mathfrak g$.  Its translation and mixed brackets are
\begin{equation}
[P_{a,b},P_{c,d}]
=\Lambda b\,\delta_{b,-d}L_{a+c}
+\Lambda a\,\delta_{a,-c}\bar L_{b+d}.
\label{eq:null-curved-translations}
\end{equation}
\begin{align}
[L_m,L_n]&=(m-n)L_{m+n},
&
[\bar L_r,\bar L_s]&=(r-s)\bar L_{r+s},
&
[L_m,\bar L_r]&=0,
\nonumber\\
[L_m,P_{a,b}]&=\left(\frac m2-a\right)P_{m+a,b},
&
[\bar L_r,P_{a,b}]&=\left(\frac r2-b\right)P_{a,r+b}.
\label{eq:null-lorentz-translation-brackets}
\end{align}
The deltas are Kronecker deltas on the discrete mode indices.  Terms outside
the global ranges have zero coefficients.  The translations commute only
at $\Lambda=0$.

Taking the leading-soft residue of the deformed graviton OPE gives
\begin{align}
P_{-,-}\lvert G^+_\Delta\rangle
&=-\frac12\lvert G^+_{\Delta+1}\rangle
+\frac{\Lambda}{2}\partial\bar\partial
 \lvert G^+_{\Delta-1}\rangle,
&
P_{-,+}\lvert G^+_\Delta\rangle
&=-\frac{\Lambda}{2}(\Delta-3)\partial
 \lvert G^+_{\Delta-1}\rangle,
\\
P_{+,-}\lvert G^+_\Delta\rangle
&=\frac{\Lambda}{2}(\Delta+1)\bar\partial
 \lvert G^+_{\Delta-1}\rangle,
&
P_{+,+}\lvert G^+_\Delta\rangle
&=-\frac{\Lambda}{2}(\Delta-3)(\Delta+1)
 \lvert G^+_{\Delta-1}\rangle.
\label{eq:null-four-origin-actions}
\end{align}
The residue calculation, including the adjacent higher-spin modes, is
given in Appendix~\ref{app:projected-null-conditions}.  The flat term raises
$\Delta$ by one, while the curvature terms lower it by one.  On descendants,
these primary formulas must be combined with the mixed commutators, as in
eqs.~\eqref{eq:null-low-hol-recursion}--\eqref{eq:null-low-anti-recursion}.

At $\Delta=3$, the $P_{-,+}$ and $P_{+,+}$ actions vanish without a
projection.  Together with Lorentz primarity, they give
\begin{equation}
\mathfrak n_R
=\operatorname{span}\{L_1,\bar L_1,P_{-,+},P_{+,+}\},
\qquad
\mathfrak n_R\lvert G^+_3\rangle=0.
\label{eq:null-right-polarization}
\end{equation}
The only nonzero brackets among these four generators are
\begin{equation}
[L_1,P_{-,+}]=P_{+,+},
\qquad
[P_{-,+},P_{+,+}]=-\frac{\Lambda}{2}\bar L_1.
\label{eq:null-right-polarization-brackets}
\end{equation}
Thus $\mathfrak n_R$ is nilpotent: repeated commutators terminate, here after
$\bar L_1$ is reached.  This generator is central in $\mathfrak n_R$.
The subalgebra defines the raising polarization used below.

The opposite translation acts as
$P_{+,-}\lvert G_3^+\rangle=2\Lambda\bar\partial\lvert G_2^+\rangle$, and
\begin{equation}
P_{-,+}\bar\partial\lvert G_2^+\rangle
=\frac12\lvert G_3^+\rangle\ne0.
\label{eq:null-global-quotient-failure}
\end{equation}
Although $\bar\partial G_2^+$ is singular under the antiholomorphic global
$\mathfrak{sl}_2$, discarding it is not a quotient preserved by the deformed
translations.  We retain this state throughout.  The projected-mode
classification and the obstruction to imposing additional translation
annihilators are recorded in Appendix~\ref{app:projected-null-conditions}.

\paragraph{Relation to the CFT$_3$ light-ray wedge.}
At $\Lambda=-1$, the maps
\eqref{eq:universal-w-map} and \eqref{eq:null-translation-identification}
give the seven generators in the direct intersection with the light-ray
seed families:
\begin{align}
L_{-1}&=\ii E_{-1},
&
L_0&=\frac12(N_0+\ii K_0),
\nonumber\\
\bar L_{-1}&=\ii E_1,
&
\bar L_0&=-\frac12(N_0-\ii K_0),
\nonumber\\
P_{-,-}&=-E_0,
&
P_{-,+}&=-\frac12(K_{-1}+\ii N_{-1}),
\nonumber\\
&&
P_{+,-}&=-\frac12(K_1-\ii N_1).
\label{eq:null-light-ray-seven-map}
\end{align}
In the Einstein-cylinder realization of
ref.~\cite{Strominger:2026cft}, the pre-existing global conformal charges
supply the remaining three generators,
\begin{equation}
L_1=-w^1_{0,1},
\qquad
\bar L_1=w^2_{1,0},
\qquad
P_{+,+}=w^{3/2}_{1/2,1/2}
=[L_1,P_{-,+}]=[\bar L_1,P_{+,-}].
\label{eq:null-light-ray-wedge-completion}
\end{equation}
These obey eq.~\eqref{eq:universal-cft3-wedge}.
We use this completed algebra through the mode-algebra isomorphism
described in Section~\ref{sec:universal-algebra}.
In the ANEC convention of that reference, the boundary mode instead satisfies
\begin{equation}
A_p=w^p_{1-p,p-2},
\qquad
[L_{-1},A_p]=[\bar L_{-1},A_p]=0.
\label{eq:null-anec-opposite-polarization}
\end{equation}
Its nonvanishing $L_1$ and $\bar L_1$ actions generate conformal descendants.
The ANEC and celestial constructions therefore furnish distinct
representations of the same low-spin algebra in opposite polarizations.

\subsection{Exact null descendants}
\label{sec:lambda-exact-null-descendants}

We distinguish an abstract highest-weight state $\lvert\Omega\rangle$ from
its celestial image $\lvert G_3^+\rangle$.  Let
$\mathfrak h=\operatorname{span}\{L_0,\bar L_0\}$ and define the freely
induced module
\begin{equation}
\mathcal M=
U(\mathfrak g)\otimes_{U(\mathfrak h\oplus\mathfrak n_R)}
\mathbb C_{5/2,\,1/2}.
\label{eq:null-induced-module}
\end{equation}
Thus $\mathfrak n_R\lvert\Omega\rangle=0$,
$L_0\lvert\Omega\rangle=\frac52\lvert\Omega\rangle$, and
$\bar L_0\lvert\Omega\rangle=\frac12\lvert\Omega\rangle$.
No further relations on lowering descendants are imposed in $\mathcal M$.
A singular vector is a nonzero descendant annihilated by $\mathfrak n_R$.

\paragraph{The cubic singular vector.}
The opposite generators obey
\begin{equation}
[L_{-1},P_{+,-}]=-P_{-,-},
\qquad
[P_{-,-},P_{+,-}]=-\frac{\Lambda}{2}\bar L_{-1},
\label{eq:null-lowering-brackets}
\end{equation}
with the other brackets among
$L_{-1},P_{-,-},P_{+,-},\bar L_{-1}$ zero.
In this order, the Poincare--Birkhoff--Witt (PBW) theorem makes
$L_{-1}P_{+,-}^2\lvert\Omega\rangle$,
$P_{-,-}P_{+,-}\lvert\Omega\rangle$ and
$\bar L_{-1}\lvert\Omega\rangle$ independent basis vectors.  They all have
weights $(5/2,3/2)$.  Consider
\begin{equation}
\lvert\chi(a,b,c)\rangle
=\left(aL_{-1}P_{+,-}^2+bP_{-,-}P_{+,-}
+c\Lambda\bar L_{-1}\right)\lvert\Omega\rangle.
\label{eq:null-chi-ansatz}
\end{equation}
Commuting the raising generators to the right gives
\begin{align}
L_1\lvert\chi(a,b,c)\rangle
&=(3a+b)P_{+,-}^2\lvert\Omega\rangle,
\nonumber\\
\bar L_1\lvert\chi(a,b,c)\rangle
&=\Lambda(b+c)\lvert\Omega\rangle,
\nonumber\\
P_{-,+}\lvert\chi(a,b,c)\rangle
&=\frac{\Lambda}{2}(3a+b)L_{-1}P_{+,-}\lvert\Omega\rangle
+\Lambda(b+c)P_{-,-}\lvert\Omega\rangle,
\nonumber\\
P_{+,+}\lvert\chi(a,b,c)\rangle
&=\Lambda\left[\frac32(a+b)+c\right]P_{+,-}\lvert\Omega\rangle.
\label{eq:null-chi-raising-actions}
\end{align}
For $\Lambda\ne0$, singularity requires $3a+b=0$ and $b+c=0$.
The solution is unique within this three-term ansatz up to normalization:
\begin{equation}
\lvert\chi\rangle
=\left(L_{-1}P_{+,-}^2-3P_{-,-}P_{+,-}
+3\Lambda\bar L_{-1}\right)\lvert\Omega\rangle,
\qquad \mathfrak n_R\lvert\chi\rangle=0.
\label{eq:null-chi-singular}
\end{equation}
PBW independence shows that $\lvert\chi\rangle\ne0$ in $\mathcal M$.

In the celestial realization, $P_{+,-}$ and $P_{-,-}$ commute with
$\bar L_{-1}$.  Combining this with the primary actions gives
\begin{align}
P_{+,-}\lvert G_3^+\rangle
&=-2\Lambda\bar L_{-1}\lvert G_2^+\rangle,
\nonumber\\
P_{+,-}^2\lvert G_3^+\rangle
&=3\Lambda^2\bar L_{-1}^2\lvert G_1^+\rangle,
\nonumber\\
P_{-,-}P_{+,-}\lvert G_3^+\rangle
&=\Lambda\bar L_{-1}\lvert G_3^+\rangle
+\Lambda^2L_{-1}\bar L_{-1}^2\lvert G_1^+\rangle.
\label{eq:null-chi-celestial-actions}
\end{align}
The $\bar L_{-1}G_3^+$ and $L_{-1}\bar L_{-1}^2G_1^+$ terms then
cancel separately:
\begin{equation}
\lvert\chi_{\rm cel}\rangle
=\left(L_{-1}P_{+,-}^2-3P_{-,-}P_{+,-}
+3\Lambda\bar L_{-1}\right)\lvert G_3^+\rangle=0.
\label{eq:null-chi-celestial-zero}
\end{equation}

\paragraph{The fifth-power singular vector.}
The same module contains a second singular vector,
\begin{equation}
\lvert\eta\rangle=P_{+,-}^5\lvert\Omega\rangle,
\qquad (h_\eta,\bar h_\eta)=(0,3).
\label{eq:null-eta-singular}
\end{equation}
For general inducing weights $(h,\bar h)$, the brackets
$[P_{-,+},P_{+,-}]=\frac{\Lambda}{2}(L_0-\bar L_0)$ and
$[L_0-\bar L_0,P_{+,-}]=-P_{+,-}$ imply
\begin{equation}
P_{-,+}P_{+,-}^r\lvert\Omega\rangle
=\frac{\Lambda r}{2}
\left(h-\bar h-\frac{r-1}{2}\right)
P_{+,-}^{r-1}\lvert\Omega\rangle.
\label{eq:null-eta-raising-action}
\end{equation}
The other three raising generators annihilate $P_{+,-}^r\lvert\Omega\rangle$.
For $h-\bar h=2$, the displayed coefficient vanishes at $r=5$.
The resulting vector is nonzero by PBW independence.

Its celestial image follows from an induction using
eq.~\eqref{eq:null-four-origin-actions}:
\begin{equation}
P_{+,-}^r\lvert G_3^+\rangle
=\left(-\frac{\Lambda}{2}\right)^r
\prod_{j=0}^{r-1}(4-j)\,
\bar L_{-1}^r\lvert G_{3-r}^+\rangle.
\label{eq:null-celestial-power}
\end{equation}
For $\Lambda\ne0$, the first four coefficients are
$-2\Lambda$, $3\Lambda^2$, $-3\Lambda^3$ and $3\Lambda^4/2$.
The fifth action vanishes:
\begin{equation}
\lvert\eta_{\rm cel}\rangle=P_{+,-}^5\lvert G_3^+\rangle=0.
\label{eq:null-eta-celestial-zero}
\end{equation}

Let $\mathcal V_{\rm cel}$ be the unprojected algebraic span of the hard states
$\lvert G_\delta^+\rangle$ and their Lorentz descendants, with all Mellin
labels needed by the shifts retained.  Write
$\mathcal V_3=U(\mathfrak g)\lvert G_3^+\rangle$ for its cyclic submodule
generated by $G_3^+$.  The highest-weight conditions define the
intertwiner
$\pi:\mathcal M\to\mathcal V_3\subseteq\mathcal V_{\rm cel}$ by
$\pi(\lvert\Omega\rangle)=\lvert G_3^+\rangle$.
The two results state
\begin{equation}
0\ne\lvert\chi\rangle,\lvert\eta\rangle\in\mathcal M,
\qquad
\pi(\lvert\chi\rangle)=\pi(\lvert\eta\rangle)=0.
\label{eq:null-abstract-concrete}
\end{equation}
The kernel of this map is invariant under $\mathfrak g$.  This is the
algebraic reason that the two null relations are consistent, in contrast
to discarding $\bar\partial G_2^+$ in
eq.~\eqref{eq:null-global-quotient-failure}.

\subsection{Casimir and shortening interpretation}
\label{sec:lambda-quadratic-casimir}

For $\Lambda\ne0$, the quadratic Casimir of the low-spin algebra, normalized
as three times the inverse-Killing Casimir, is
\begin{align}
\mathcal C_2={}&
L_0^2-\frac12\{L_{-1},L_1\}
+\bar L_0^2-\frac12\{\bar L_{-1},\bar L_1\}
\nonumber\\
&+\frac1{\Lambda}\left(
\{P_{-,+},P_{+,-}\}-\{P_{-,-},P_{+,+}\}\right).
\label{eq:null-quadratic-casimir}
\end{align}
It satisfies
\begin{equation}
[\mathcal C_2,L_a]=[\mathcal C_2,\bar L_b]
=[\mathcal C_2,P_{c,d}]=0.
\label{eq:null-casimir-centrality}
\end{equation}
The descendant action in Appendix~\ref{app:celestial-module-checks}
gives
\begin{equation}
\mathcal C_2\lvert G^+_\Delta\rangle
=3\lvert G^+_\Delta\rangle.
\label{eq:null-casimir-eigenvalue}
\end{equation}
The two translation anticommutators contain shifted descendants, but these
cancel between the two terms.  The remaining translation contribution
cancels the $\Delta$ dependence of the Lorentz Casimir.  The individual
terms are displayed in
eqs.~\eqref{eq:null-translation-anticommutators}--\eqref{eq:null-lorentz-casimir-piece}.

In the Cartesian convention of ref.~\cite{Taylor:2023ajb},
\begin{equation}
C_2^{\rm std}:=\frac12 M_{AB}M^{AB}=2\mathcal C_2,
\qquad
C_2^{\rm std}(E_0^{\rm AdS},s)=E_0^{\rm AdS}(E_0^{\rm AdS}-3)+s(s+1),
\label{eq:null-standard-ads-casimir}
\end{equation}
Here $E_0^{\rm AdS}$ and $s$ are the lowest energy and spin in the AdS$_4$
positive-energy convention \cite{Karch:2000ct,Dolan:2005wy}.
The superscript distinguishes this energy label from the light-ray
zero mode $E_0$ in eq.~\eqref{eq:null-light-ray-seven-map}.
On a low-spin highest-weight state,
$E_0^{\rm AdS}=h+\bar h$ and $s=h-\bar h$; the general evaluation is given in
eq.~\eqref{eq:null-low-spin-highest-weight-casimir}.  The state
$G_3^+$ therefore has $(E_0^{\rm AdS},s)=(3,2)$ and $C_2^{\rm std}=6$, the massless
spin-two value.  Other $G_\Delta^+$ have the same Casimir but are not
separate low-spin highest-weight states with $E_0^{\rm AdS}=\Delta$.
The deformed translations connect their Mellin labels inside the
dimension-completed representation.  Centrality extends the eigenvalue to
their descendants.

The null vectors identify the shortening more precisely than the Casimir
alone.  Under the energy grading $L_0+\bar L_0$,
$L_{-1}$, $P_{-,-}$ and $\bar L_{-1}$ raise energy by one, whereas
$P_{+,-}$ preserves energy and lowers the spin weight.
Consequently,
\begin{equation}
(E_\Omega,s_\Omega)=(3,2),
\qquad
(E_\chi,s_\chi)=(4,1).
\label{eq:null-shortening-weights}
\end{equation}
The cubic polynomial in $\lvert\chi\rangle$ is an energy-level-one
descendant.  The relation $P_{+,-}^5\lvert\Omega\rangle=0$ gives the
finite spin-two ladder.  Moreover,
\begin{equation}
P_{+,-}^3\lvert\chi\rangle
=L_{-1}P_{+,-}^5\lvert\Omega\rangle,
\label{eq:null-finite-spin-descendant}
\end{equation}
so after imposing the finite-spin relation, $\lvert\chi\rangle$ generates
a spin-one multiplet.

Let $\mathcal V(E,s)$ denote the generalized Verma module induced from
finite spin $s$.  The massless spin-two shortening has the sequence
\begin{equation}
0\longrightarrow\mathcal V(4,1)
\longrightarrow\mathcal V(3,2)
\longrightarrow D(3,2)\longrightarrow0,
\label{eq:null-shortening-sequence}
\end{equation}
where the first map is generated by $\lvert\chi\rangle$ and $D(3,2)$
denotes the irreducible quotient.  The explicit descendant and norm agree
with the spin-two specialization of Dolan's construction
\cite[Eqs.~(3.42), (C.5)--(C.6)]{Dolan:2005wy};
the generator map and norm calculation are in
Appendix~\ref{app:celestial-shortening-checks}.
The celestial zeros \eqref{eq:null-chi-celestial-zero} and
\eqref{eq:null-eta-celestial-zero} are algebraic state identities.
The AdS comparison identifies the shortening of the generated low-spin
highest-weight sector, with the contravariant norm evaluated using the
adjoint specified in Appendix~\ref{app:celestial-shortening-checks}.
The Casimir \eqref{eq:null-quadratic-casimir} itself has no nonsingular
flat contraction because its Killing form degenerates at $\Lambda=0$.

\subsection{Transport and differential--difference equations}
\label{sec:lambda-null-decoupling-equations}

Consider the dimension-completed correlator
\begin{equation}
\mathcal A_{n+1}
=\left\langle G_3^+(z_0,\bar z_0)
\prod_{i=1}^nG_{\Delta_i}^+(z_i,\bar z_i)\right\rangle
\label{eq:null-correlator}
\end{equation}
with an invariant vacuum and the low-spin Ward identities.
All external sums below run over $i=1,\ldots,n$.
Set $z_{i0}=z_i-z_0$, $\bar z_{i0}=\bar z_i-\bar z_0$,
$\partial_i=\partial/\partial z_i$ and
$\bar\partial_i=\partial/\partial\bar z_i$.
Let $\mathsf E_i^\pm$ shift the $i$th Mellin label:
\begin{equation}
\mathsf E_i^\pm
\mathcal A(\ldots,\Delta_i,\ldots)
=\mathcal A(\ldots,\Delta_i\pm1,\ldots),
\label{eq:null-mellin-shift-operators}
\end{equation}
The four origin-primary actions are represented by
\begin{align}
d_{-,-}^{(i)}&=-\frac12\mathsf E_i^+
+\frac{\Lambda}{2}\partial_i\bar\partial_i\mathsf E_i^-,
\nonumber\\
d_{-,+}^{(i)}&=-\frac{\Lambda}{2}(\Delta_i-3)
\partial_i\mathsf E_i^-,
\nonumber\\
d_{+,-}^{(i)}&=\frac{\Lambda}{2}(\Delta_i+1)
\bar\partial_i\mathsf E_i^-,
\nonumber\\
d_{+,+}^{(i)}&=-\frac{\Lambda}{2}(\Delta_i-3)(\Delta_i+1)
\mathsf E_i^-.
\label{eq:null-origin-ward-operators}
\end{align}
As in flat celestial null-decoupling equations
\cite{Banerjee:2020zlg,Banerjee:2023rni}, Mellin shifts make these
differential--difference operators.  Curvature supplies the downward shifts
and the noncommuting translation actions.

\paragraph{Transported generators and highest-weight equations.}
To move the distinguished state $G^+_3$ from the origin to
$(z_0,\bar z_0)$, define
\[
A_0=z_0L_{-1}-\bar z_0\bar L_{-1},
\qquad U_0=e^{A_0},
\qquad
\lvert G^+_3(z_0,\bar z_0)\rangle
=U_0\lvert G^+_3(0)\rangle.
\]
The minus sign implements the residue convention
$\bar L_{-1}G^+_\Delta=-\bar\partial G^+_\Delta$.
If $Q\lvert G^+_3(0)\rangle=0$, its transported annihilator is
$U_0QU_0^{-1}$, since
\[
(U_0QU_0^{-1})\lvert G^+_3(z_0,\bar z_0)\rangle
=U_0Q\lvert G^+_3(0)\rangle=0.
\]
For $Q=P_{-,+},P_{+,+}$ in
eq.~\eqref{eq:null-right-polarization}, the adjoint expansion is
\[
U_0QU_0^{-1}
=Q+[A_0,Q]+\frac{1}{2!}[A_0,[A_0,Q]]+\cdots.
\]
Using the Lorentz--translation brackets
\eqref{eq:null-lorentz-translation-brackets}, we obtain
\begin{align*}
[A_0,P_{-,+}]&=\bar z_0P_{-,-},
\\
[A_0,P_{+,+}]&=-z_0P_{-,+}+\bar z_0P_{+,-},
\\
[A_0,[A_0,P_{+,+}]]&=-2z_0\bar z_0P_{-,-}.
\end{align*}
Since $[A_0,P_{-,-}]=0$, the two series terminate after the first and
second commutators, respectively.  The factor $1/2!$ leaves the mixed term
$-z_0\bar z_0P_{-,-}$ in the second annihilator.  Thus
$\mathcal Q_1=U_0P_{-,+}U_0^{-1}$ and
$\mathcal Q_2=U_0P_{+,+}U_0^{-1}$ are exactly
\begin{align}
\mathcal Q_1(z_0,\bar z_0)
&=P_{-,+}+\bar z_0P_{-,-},
\nonumber\\
\mathcal Q_2(z_0,\bar z_0)
&=P_{+,+}-z_0P_{-,+}+\bar z_0P_{+,-}
-z_0\bar z_0P_{-,-}.
\label{eq:null-transported-annihilators}
\end{align}
Both annihilate $\lvert G^+_3(z_0,\bar z_0)\rangle$.
The finite adjoint series is exact for arbitrary
$z_0,\bar z_0$ and $\Lambda$.
Conjugating the second bracket in
eq.~\eqref{eq:null-right-polarization-brackets} by $U_0$ gives their
commutator, which closes on the transported global raising mode,
\begin{equation}
[\mathcal Q_1,\mathcal Q_2]
=-\frac{\Lambda}{2}
\left(\bar L_1+2\bar z_0\bar L_0+\bar z_0^2\bar L_{-1}\right).
\label{eq:null-ward-integrability}
\end{equation}

Conjugation also gives
\begin{equation}
\widetilde P_{+,-}
=U_0P_{+,-}U_0^{-1}=P_{+,-}-z_0P_{-,-},
\label{eq:null-transported-lowerer}
\end{equation}
while $L_{-1}$, $\bar L_{-1}$ and $P_{-,-}$ are unchanged.
On an external leg, the combination
$U_i^{-1}\widetilde P_{+,-}U_i$, with
$U_i=e^{z_iL_{-1}-\bar z_i\bar L_{-1}}$, is
$P_{+,-}+z_{i0}P_{-,-}$.  Its differential action is therefore
\begin{equation}
d_{+,-}^{(i)}+z_{i0}d_{-,-}^{(i)}
=-\frac{z_{i0}}2\mathsf E_i^+
+\frac{\Lambda}{2}(\Delta_i+1+z_{i0}\partial_i)
\bar\partial_i\mathsf E_i^-.
\label{eq:null-transported-leg-action}
\end{equation}

Moving $\mathcal Q_1$ and $\mathcal Q_2$ to the external legs gives the two
highest-weight Ward equations
\begin{align}
\sum_{i=1}^{n}
\left(d_{-,+}^{(i)}-\bar z_{i0}d_{-,-}^{(i)}\right)
\mathcal A_{n+1}=0,
\label{eq:null-first-ward-equation}\\
\sum_{i=1}^{n}
\left(d_{+,+}^{(i)}+z_{i0}d_{-,+}^{(i)}
-\bar z_{i0}d_{+,-}^{(i)}
-z_{i0}\bar z_{i0}d_{-,-}^{(i)}\right)
\mathcal A_{n+1}=0.
\label{eq:null-second-ward-equation}
\end{align}
Their compatibility with the antiholomorphic global Ward identity follows
from eq.~\eqref{eq:null-ward-integrability}.

\paragraph{Ordered transfer of null descendants.}
At the marked point the two exact null relations are
\begin{align}
\left[L_{-1}(\widetilde P_{+,-})^2
-3P_{-,-}\widetilde P_{+,-}
+3\Lambda\bar L_{-1}\right]
\lvert G_3^+(z_0,\bar z_0)\rangle&=0,
\nonumber\\
(\widetilde P_{+,-})^5
\lvert G_3^+(z_0,\bar z_0)\rangle&=0.
\label{eq:null-transported-descendants}
\end{align}
To convert products into differential operators, write
$\mathcal F(w)=\langle w\prod_iG_{\Delta_i}^+\rangle$.
For a transported generator $Q$, vacuum invariance gives
\begin{equation}
\mathcal F(Qw)=-\sum_i d_Q^{(i)}\mathcal F(w).
\label{eq:null-product-ward-recursion}
\end{equation}
The other insertions remain primaries.  Removing the outermost generator
repeatedly therefore yields
\begin{equation}
\mathcal F(Q_1\cdots Q_k w)
=(-1)^k\left(\sum_i d_{Q_1}^{(i)}\right)\cdots
\left(\sum_j d_{Q_k}^{(j)}\right)\mathcal F(w).
\label{eq:null-ordered-ward-transfer}
\end{equation}
The order of the differential operators is the order of the word.
The abstract transfer of charges reverses products, but converting their
actions on primaries to differential operators introduces a second reversal:
$d_{[Q,R]}=-[d_Q,d_R]$.  For example,
$[z\partial+h,\partial]=-\partial$, whereas
$[L_0,L_{-1}]=L_{-1}$.

All products below are ordered compositions.  Derivatives act on
coordinate-dependent coefficients, and
\begin{equation}
\mathsf E_i^\pm f(\Delta_i)
=f(\Delta_i\pm1)\mathsf E_i^\pm.
\label{eq:null-shift-ordering}
\end{equation}
For example,
\begin{equation}
\bigl(d_{+,-}^{(i)}\bigr)^2
=\frac{\Lambda^2}{4}(\Delta_i+1)\Delta_i
\bar\partial_i^2(\mathsf E_i^-)^2.
\label{eq:null-shift-square-example}
\end{equation}
The first Mellin shift changes the coefficient of the second factor from
$\Delta_i+1$ to $\Delta_i$.

The three terms of the cubic null have external Ward images
\begin{align}
L_{-1}(\widetilde P_{+,-})^2
&\longmapsto-\sum_i\partial_i
\left(\sum_j[d_{+,-}^{(j)}+z_{j0}d_{-,-}^{(j)}]\right)^2,
\nonumber\\
P_{-,-}\widetilde P_{+,-}
&\longmapsto+\sum_{i,j}d_{-,-}^{(i)}
[d_{+,-}^{(j)}+z_{j0}d_{-,-}^{(j)}],
\nonumber\\
\bar L_{-1}&\longmapsto+\sum_i\bar\partial_i.
\label{eq:null-three-ward-images}
\end{align}
The last sign combines $\bar L_{-1}=-\bar\partial$ with the Ward-transfer
minus sign.  Define
\begin{equation}
\begin{aligned}
\mathcal C_\chi={}&\sum_i\partial_i
\left(\sum_j[d_{+,-}^{(j)}+z_{j0}d_{-,-}^{(j)}]\right)^2\\
&+3\sum_{i,j}d_{-,-}^{(i)}
[d_{+,-}^{(j)}+z_{j0}d_{-,-}^{(j)}]
-3\Lambda\sum_i\bar\partial_i .
\end{aligned}
\label{eq:null-cubic-ward-operator}
\end{equation}
The new null-decoupling equations are
\begin{align}
\mathcal C_\chi\mathcal A_{n+1}=0,
\label{eq:null-cubic-ward-equation}\\
\left(\sum_i[d_{+,-}^{(i)}+z_{i0}d_{-,-}^{(i)}]\right)^5
\mathcal A_{n+1}=0.
\label{eq:null-fifth-ward-equation}
\end{align}
The double sum includes $i=j$, and the powers act on the full sums,
including cross-leg terms.  The leading derivatives in
eq.~\eqref{eq:null-cubic-ward-operator} also differentiate the coefficients
inside the square.  An equivalent right-ordered formula, the one-leg
square, and a check using derivatives at the marked insertion are given in
Appendix~\ref{app:celestial-ward-ordering}.

The two null relations hold in the concrete celestial representation.
For one external leg, the cubic Ward operator factorizes through the
highest-weight and global Ward operators, as shown explicitly in
Appendix~\ref{app:celestial-ward-independence}.

\section{Concluding remarks}
\label{sec:concluding-remarks}

The scalar and fermion theories provide two microscopic realizations of the
same three-family sector of $\mathcal{L}_{\Lambda}w_{1+\infty}$ at
$\Lambda=-1$.  Their local stress tensors and canonical statistics differ,
but their light-ray moments reduce to identical one-particle differential
operators.  The leading critical $O(N)$ construction is the corresponding
flavor sum and closes on the fundamental-field Fock sector and its singlets.

At finite $N$, we use the light-ray algebra under the closure and domain
assumptions stated in Section~\ref{sec:universal-algebra}.
Prescribed iterated zero-frequency limits of the detector kernel are
consistent with its pairwise coefficients.  The resulting Ward-algebra
matrix elements in eq.~\eqref{eq:on-NLO-probes} carry the expected
$1/N$ corrections through the external-state dimensions.

The celestial calculation identifies an exact low-spin highest-weight
state $G_3^+$ and two null descendants in its unprojected realization.
The cubic descendant $\lvert\chi\rangle$ and the fifth-power descendant
$\lvert\eta\rangle$ are nonzero singular vectors in the freely induced
module, but their celestial images vanish.  The first gives the
energy-level-one spin-one shortening of the massless spin-two sector; the
second supplies its finite-spin relation.  The nonzero-$\Lambda$ quadratic
Casimir has eigenvalue three throughout the dimension-completed hard
family.

The ordered Ward images of the two descendants give
eqs.~\eqref{eq:null-cubic-ward-equation} and
\eqref{eq:null-fifth-ward-equation}, supplementing the highest-weight
equations \eqref{eq:null-first-ward-equation} and
\eqref{eq:null-second-ward-equation}.  Their derivation retains the action
of Mellin shifts and coordinate derivatives on operator coefficients.
A next step is to characterize the multi-leg solution space and the
independent constraints within this differential--difference system.

For the interacting light-ray problem, a direct approach is to evaluate
the required moments with finite null-time and sheet regulators, then
control their removal in the matrix elements of interest.  This requires bounds on the
$TTT$, $TT\sigma$, and higher-singlet contributions before taking their
isolated detector limit.  Nested commutators require additional control
because terms suppressed in one pair can contribute to a second commutator.
The light-ray OPE and frequency-resolved frameworks
\cite{Kologlu:2019mfz,Korchemsky:2021htm,Hartman:2023qdn} provide tools for
these questions.  Extending the oscillator construction to the collective
$\sigma$ sector is another natural problem.

A concrete string-theory setting for the interacting light-ray problem is
Aharony--Bergman--Jafferis--Maldacena (ABJM) theory.
At large $N$, the level-$k$ theory is dual to M-theory on
$\mathrm{AdS}_4\times S^7/\mathbb Z_k$, while its 't~Hooft limit is described
by type IIA string theory on $\mathrm{AdS}_4\times\mathbb{CP}^3$
\cite{Aharony:2008ug}.  The averaged null energy condition (ANEC) construction of
ref.~\cite{Strominger:2026cft} then supplies the corresponding
$\mathcal{L}_{-1}w_{1+\infty}$ stress-tensor light-ray action in ABJM, subject
to the global and domain assumptions of that construction.  ABJM is therefore a
concrete strongly coupled theory in which to seek a bulk realization of these
generators and their string- or M-theory interpretation.

Sen showed that an asymptotically flat string theory fixed at
one point of its moduli space at infinity can contain an arbitrarily large
region approximating another point \cite{Sen:2025decorating}.  Can regulated
light-ray charges be defined in such finite-region constructions?
This question requires a definition of finite-patch charges and their
boundary conditions.

\section*{Acknowledgements}
BZ is supported by the Fundamental Research Funds for the Central Universities
(010-63263123). We thank OpenAI Codex (GPT 5.6) for assistance with drafting,
editing, consistency checks, and manuscript organization.
The author is fully responsible for the scientific ideas, analytical strategy,
derivations, conclusions, and manuscript preparation.

\appendix

\section{Canonical reductions and normalization}
\label{app:canonical-details}

This appendix records the canonical reductions and local contact terms used
for both free-field realizations, together with the collective-field
normalization used in the critical model.  The restriction of the scalar to
the null sheet is normalized as
\begin{equation}
\phi(u,y)=\int_0^\infty\frac{\dd p}{2\pi}
\int_{-\infty}^{+\infty}\frac{\dd q}{2\pi}\frac{1}{\sqrt{2p}}
\left[a(p,q)e^{-\ii pu+\ii qy}
+a^\dagger(p,q)e^{\ii pu-\ii qy}\right],
\label{eq:scalar-mode-expansion}
\end{equation}
where $p>0$ is the longitudinal light-front momentum, $q$ is the transverse
momentum, and $a(p,q)$ is the scalar annihilation operator.  Its adjoint is
$a^\dagger(p,q)$, and the canonical commutator is
\begin{equation}
[a(p,q),a^\dagger(p',q')]
=(2\pi)^2\delta(p-p')\delta(q-q').
\label{eq:scalar-canonical-commutator}
\end{equation}
Because $p,p'>0$, integration over the complete null generator removes the
pair-creation and pair-annihilation terms.  The number-conserving terms use
\begin{equation}
\int\dd u\,e^{\ii(p-p')u}=2\pi\delta(p-p'),
\qquad
\int\dd u\,u e^{\ii(p-p')u}=-2\pi\ii\,\delta'(p-p'),
\label{eq:scalar-null-Fourier-identities}
\end{equation}
where $\delta'$ denotes the distributional derivative of the delta function.
These identities produce the kernels in eq.~\eqref{eq:scalar-EKN-modes} and, in
particular, the contact term in eq.~\eqref{eq:scalar-delta-prime}.

Writing
$a(p,y)=\int\dd q\,e^{\ii qy}a(p,q)/(2\pi)$ gives
$[a(p,y),a^\dagger(p',y')]=2\pi\delta(p-p')\delta(y-y')$.
The mixed component reduces to
\begin{align}
\cN(y)
&=\int_0^\infty\frac{\dd p}{2\pi}\,
\frac{\ii}{2}\normord{a^\dagger(p,y)
\overleftrightarrow{\partial_y}a(p,y)},
\nonumber\\
N[Y]
&=\int_0^\infty\frac{\dd p}{2\pi}\int\dd y\,
\normord{a^\dagger
\ii\left(Y\partial_y+\frac12Y'\right)a}.
\label{eq:scalar-N-line-kernel}
\end{align}
For a test wavefunction $f(p)$, the first null moment instead uses
\begin{equation}
\int_0^\infty\dd p'\,\delta'(p-p')\sqrt{pp'}\,f(p')
=\left(p\partial_p+\frac12\right)f(p),
\label{eq:scalar-delta-prime}
\end{equation}
which fixes the $1/2$ term by the stress-tensor matrix element rather than by
an ordering convention.

The improvement terms are complete null derivatives.  Under the endpoint
conditions stated in Section~\ref{sec:free-scalar},
\begin{align}
\int\dd u\,\partial_u^2\phi^2
&=\left[\partial_u\phi^2\right]_{-\infty}^{+\infty}=0,
\nonumber\\
\int\dd u\,u\partial_u^2\phi^2
&=\left[u\partial_u\phi^2-\phi^2\right]_{-\infty}^{+\infty}=0,
\nonumber\\
\int\dd u\,\partial_u\partial_y\phi^2
&=\partial_y\left[\phi^2\right]_{-\infty}^{+\infty}=0.
\label{eq:scalar-improvement-boundaries}
\end{align}
Equivalently, one may first smear in $u$, compute all distributional terms,
and then remove the regulator on this wave-packet domain.

For the fermion, the symmetric action differs from the standard Dirac action
by the total derivative
\begin{equation}
\frac{\ii}{2}\bar\psi\gamma^\mu
\overleftrightarrow{\partial_\mu}\psi
=\ii\bar\psi\gamma^\mu\partial_\mu\psi
-\frac{\ii}{2}\partial_\mu(\bar\psi\gamma^\mu\psi).
\label{eq:fermion-action-boundary}
\end{equation}
Our Euclidean normalization is
\begin{equation}
\langle\psi_\alpha(x)\bar\psi_\beta(0)\rangle_{\mathrm E}
=\frac{(\gamma_E\cdot x)_{\alpha\beta}}{4\pi|x|^3},
\qquad
\{\gamma_E^i,\gamma_E^j\}=2\delta^{ij},
\label{eq:fermion-propagator}
\end{equation}
which fixes $\Delta_\psi=1$ and solves the Euclidean Dirac equation away from
the coincident point.  The Lorentzian equations of motion similarly give
\begin{equation}
T^\mu{}_{\mu}=\frac{\ii}{2}
\left[\bar\psi\gamma^\mu\partial_\mu\psi
-(\partial_\mu\bar\psi)\gamma^\mu\psi\right]=0,
\qquad
\partial^\mu T_{\mu\nu}=0.
\label{eq:fermion-stress-checks}
\end{equation}

Only one complex spinor component is independent on the light front.  A
particle mode with $p>0$ and transverse momentum $q$ may be chosen as
\begin{equation}
u(p,q)=\frac{1}{2\sqrt2p}
\begin{pmatrix}-\ii(2p+\ii q)\\ 2p-\ii q\end{pmatrix},
\qquad
e^{-\ii p x^+-\ii q^2x^-/(4p)+\ii qy}.
\label{eq:fermion-light-front-spinor}
\end{equation}
For two modes with transverse momenta $q$ and $r$, it solves the massless
Dirac equation and obeys
\begin{equation}
\bar u(p,q)\gamma_+u(p,r)=1,
\qquad
\bar u(p,q)\gamma_yu(p,r)=-\frac{q+r}{2p}.
\label{eq:fermion-light-front-bilinears}
\end{equation}
The antiparticle contractions are identical.  Let $c(p,q)$ and $d(p,q)$
denote the particle and antiparticle annihilation operators, and write
$s\in\{c,d\}$.  The two terms in
$T_{+y}^\psi$ each contribute $-(q+r)/4$, so
\begin{align}
\int\dd u\,T_{+y}^{\psi}
&=\sum_{s=c,d}\int_0^\infty\frac{\dd p}{2\pi}
\int\frac{\dd q\,\dd r}{(2\pi)^2}
\left(-\frac{q+r}{2}\right)e^{\ii(r-q)y}
\normord{s^\dagger(p,q)s(p,r)}
\nonumber\\
&=\sum_{s=c,d}\int_0^\infty\frac{\dd p}{2\pi}\,
\frac{\ii}{2}\normord{s^\dagger(p,y)
\overleftrightarrow{\partial_y}s(p,y)}.
\label{eq:fermion-N-line-density}
\end{align}

With the normalization in eq.~\eqref{eq:scalar-cylinder-map}, the nonzero
circle anticommutators are
\begin{equation}
\{c(p,\varphi),c^\dagger(p',\varphi')\}
=\{d(p,\varphi),d^\dagger(p',\varphi')\}
=2\pi\delta(p-p')\delta_{2\pi}(\varphi-\varphi').
\label{eq:fermion-light-front-CAR}
\end{equation}
The periodic delta function has the normalization specified below
eq.~\eqref{eq:scalar-cylinder-map}.
The number-conserving null-stress density relative to $\dd\varphi$ is
\begin{align}
&J(\varphi)T_{++}^{\psi}(u,0,y(\varphi))\big|_{\rm nc}
\nonumber\\
&\qquad=\frac{1}{2\pi}\sum_{s=c,d}\int_0^\infty\frac{\dd p\,\dd p'}{(2\pi)^2}
\normord{s^\dagger(p,\varphi)
\frac{p+p'}{2}e^{\ii(p-p')u}s(p',\varphi)},
\label{eq:fermion-light-front-kernel}
\end{align}
where $J=\dd y/\dd\varphi$ as in eq.~\eqref{eq:line-cylinder-densities}.
Pair-creation terms have support only at $p+p'=0$ and vanish on the regulated
$p,p'>0$ domain.  The zeroth moment gives the circle density
$\cE_{\psi,\rm circ}(\varphi)=J(\varphi)\cE_\psi(y(\varphi))$,
\begin{equation}
\cE_{\psi,\rm circ}(\varphi)=\frac{1}{2\pi}\int_0^\infty\frac{\dd p}{2\pi}\,p
\left[\normord{c^\dagger c}+\normord{d^\dagger d}\right],
\label{eq:fermion-ANEC-positive}
\end{equation}
which is a positive number-operator integral.  Smearing with
$\dd\varphi\,e^{\ii k\varphi}$ gives $E_k^\psi$ in
eq.~\eqref{eq:fermion-EKN-modes}.  The first moment differentiates both the
wave function and the stress-tensor prefactor,
\begin{equation}
\left.\partial_{p'}\left[\frac{p+p'}{2}f(p')\right]\right|_{p'=p}
=\left(p\partial_p+\frac12\right)f(p),
\label{eq:fermion-K-contact}
\end{equation}
while smearing eq.~\eqref{eq:fermion-N-line-density} and integrating by parts
gives
\begin{equation}
\widehat N_\psi[Y]
=\ii\left(Y\partial_y+\frac12Y'\right).
\label{eq:fermion-N-smeared-kernel}
\end{equation}
These are the three reduced kernels summarized in
eq.~\eqref{eq:fermion-reduced-kernels}.

For the fermion, the elementary CAR identity is
\begin{equation}
[s_i^\dagger s_j,s_r^\dagger s_t]
=\delta_{jr}s_i^\dagger s_t-\delta_{ti}s_r^\dagger s_j,
\qquad s\in\{c,d\},
\label{eq:fermion-elementary-bilinear}
\end{equation}
where the composite indices $i,j,r,t$ include the momentum and angular
arguments, and $\delta_{ij}$ is the identity kernel for $\dd\mu$.
After contraction with two one-particle kernels, this identity
implies eq.~\eqref{eq:fermion-bilinear-lift}.  The two species commute because
their bilinears are even.  The potentially local term can be displayed before
smearing by setting
$\rho_s(p,\varphi)=s^\dagger(p,\varphi)s(p,\varphi)$:
\begin{align}
[\rho_s(p,\varphi),\rho_s(q,\varphi')]
&=2\pi\delta(p-q)\delta_{2\pi}(\varphi-\varphi')
\nonumber\\
&\quad\times\left[
s^\dagger(p,\varphi)s(q,\varphi')
-s^\dagger(q,\varphi')s(p,\varphi)\right].
\label{eq:fermion-density-contact}
\end{align}
The delta functions identify both arguments, so the expression in brackets
vanishes as an operator-valued distribution.  Hence
$[\cE_{\psi,\rm circ}(\varphi),\cE_{\psi,\rm circ}(\varphi')]=0$ without an additional central term
on the normal-ordered wave-packet domain.

The collective-field normalization in
eq.~\eqref{eq:on-sigma-normalization} follows from the scalar bubble; the
corresponding quadratic effective action $S_{\rm eff}^{(2)}$ is given below.
In this Euclidean integral, $p$ denotes the external three-momentum and $q$
the loop three-momentum:
\begin{equation}
B(p)=\int\frac{\dd^3q}{(2\pi)^3}\frac{1}{q^2(q+p)^2}
=\frac{1}{8|p|},
\qquad
S_{\rm eff}^{(2)}=-\frac14\int\frac{\dd^3p}{(2\pi)^3}
\sigma(p)B(p)\sigma(-p).
\label{eq:on-sigma-bubble}
\end{equation}
On the standard large-$N$ contour, the inverse kernel is
$\langle\sigma(p)\sigma(-p)\rangle=-16|p|$.  Together with
$\int\dd^3p\,e^{\ii p\cdot x}|p|/(2\pi)^3=-1/(\pi^2|x|^4)$, this gives the
positive position-space normalization quoted in the main text.

A second representative kernel commutator, acting on a test wavefunction
$f(p,z)$, is
\begin{equation}
[\widehat N_k,\widehat E_l]f
=-z^{k+1}\partial_z(z^lpf)
+z^lp\,z^{k+1}\partial_z f
=-l\widehat E_{k+l}f.
\label{eq:scalar-NE-derivation}
\end{equation}
The half-density connection cancels.  A component Jacobi check
that probes three different families is
\begin{align}
&[N_k,[K_l,E_m]]+[K_l,[E_m,N_k]]+[E_m,[N_k,K_l]]
\nonumber\\
&\hspace{15mm}
=\ii(l+m)E_{k+l+m}-\ii mE_{k+l+m}-\ii lE_{k+l+m}=0.
\label{eq:scalar-Jacobi-example}
\end{align}

Substitution of eq.~\eqref{eq:universal-w-map} into the abstract
bracket \eqref{eq:Lambda-w-bracket} at $\Lambda=-1$ gives all six independent
family brackets,
\begin{align}
[W_E(k),W_E(l)]&=0,
&
[W_E(k),W_+(l)]&=\frac{1-k}{2}W_E(k+l),
\nonumber\\
[W_E(k),W_-(l)]&=-\frac{1+k}{2}W_E(k+l),
&
[W_+(k),W_+(l)]&=\frac{l-k}{2}W_+(k+l),
\nonumber\\
[W_+(k),W_-(l)]
&=-\frac{k}{2}W_+(k+l)+\frac{l}{2}W_-(k+l),
&
[W_-(k),W_-(l)]&=\frac{l-k}{2}W_-(k+l).
\label{eq:scalar-w-family-brackets}
\end{align}
Together with antisymmetry, these relations satisfy the Jacobi identity.  At
$\Lambda=0$ the second line of eq.~\eqref{eq:Lambda-w-bracket} vanishes.

\section{Regulator and smearing prescriptions}
\label{app:regulators}

The algebra on the transverse line is first defined with compactly supported
test functions.  For example, translating the local C\'ordova--Shao
commutator \cite{Cordova:2018ygx} to our conventions gives
\begin{equation}
[\cN(y_1),\cE(y_2)]
=-\ii\delta(y_{12})\partial_{y_2}\cE(y_2)
+\ii\partial_{y_1}\delta(y_{12})\cE(y_2).
\label{eq:reg-local-NE}
\end{equation}
Smearing with $Y(y_1)f(y_2)$ and integrating by parts gives
\begin{align}
&-\ii\int\dd y\,Yf\,\partial_y\cE
-\ii\int\dd y\,Y'f\,\cE
=\ii\int\dd y\,Yf'\cE,
\label{eq:reg-smeared-NE}
\end{align}
which is the fourth relation in
eq.~\eqref{eq:universal-smeared-algebra}.  Compact support removes transverse
boundary terms.  Choose $\chi\in C_c^\infty(\mathbb R)$ with $\chi=1$ near
the origin and define
\begin{equation}
\chi_R(y)=\chi(y/R),
\qquad
h_R(y)=y\chi_R(y).
\label{eq:on-regulated-test-functions}
\end{equation}
Constant and linear smearings are defined only after taking matrix elements,
by this cutoff sequence.  The $|y|^{-4}$ falloff in
eq.~\eqref{eq:on-one-ray-profile} justifies the limit by dominated
convergence.

The line and algebraic-circle prescriptions are related by the pullback of
scalar and vector smearings.  For
$y=\tan(\varphi/2)$, the density weights in
eq.~\eqref{eq:line-cylinder-densities} follow by requiring the three smeared
charges to be invariant.  In particular, a circle vector field
$Y^\varphi=e^{\ii k\varphi}$ corresponds to the line component
$Y^y=J e^{\ii k\varphi}$, whereas the scalar smearings of $\cE$ and $\cK$
are pulled back without this vector Jacobian.  The oscillator half-density
transformation in eq.~\eqref{eq:scalar-cylinder-map} then produces the
$-k/2$ term in $N_k$.  The chart omits $\varphi=\pi$, corresponding to
$|y|=\infty$.  The global
circle modes are therefore understood on wave packets with sufficient
transverse decay, or as limits of compactly supported transverse smearings,
so that endpoint terms vanish.  On this domain the charge pullback preserves
the mode algebra.

Along the null generator, all free-field charges are evaluated on wave-packet
states for which the fields and the weighted derivatives appearing in
eq.~\eqref{eq:scalar-improvement-boundaries} vanish at $u=\pm\infty$.  One may
equivalently insert a smooth null cutoff, perform the distributional
calculation, and remove the cutoff after the matrix element is formed.  The
$p=0$ mode is excluded until this limit is taken.

For interacting frequency flow, the two null sheets are separated by
$\epsilon>0$.  The kernel check uses the opposite-frequency region
$\omega_1=x>0$, $\omega_2=-s<0$.  In this region, the scalar frequency
kernel of ref.~\cite{Korchemsky:2021htm} specializes in three dimensions to
\begin{equation}
f_\phi(\omega_1,\omega_2)
=-\omega_2^{-3}\,
{}_2F_1\!\left(3,\frac52;5;1+\frac{\omega_1}{\omega_2}\right).
\label{eq:on-scalar-frequency-structure}
\end{equation}
The quadratic hypergeometric identity gives the real expression
\begin{equation}
F(x,s):=x^2s^2f_\phi(x,-s)
=\frac{16x^{3/2}s^{3/2}}{(\sqrt{x}+\sqrt{s})^4},
\qquad x,s>0.
\label{eq:on-scalar-frequency-closed}
\end{equation}
At fixed $x>0$,
\begin{equation}
F(x,s)
=16\frac{s^{3/2}}{\sqrt{x}}-64\frac{s^2}{x}+O(s^{5/2}).
\label{eq:on-scalar-frequency-axis-expansion}
\end{equation}
Define the iterated kernel operations by
\begin{equation}
\mathcal M_{ab}H
:=\lim_{x\to0^+}\lim_{s\to0^+}
\partial_x^a\partial_s^bH(x,s),
\qquad a,b\in\{0,1\}.
\label{eq:on-iterated-kernel-prescription}
\end{equation}
Derivatives are taken before either limit.  The inner limit keeps $x>0$
fixed.  Equation~\eqref{eq:on-scalar-frequency-axis-expansion} gives
$\mathcal M_{ab}F=0$ for all four choices, including those used in
eq.~\eqref{eq:on-scalar-frequency-moments}.  This is an iterated-axis
limit.  Its dependence on the approach to the origin is explicit:
\begin{equation}
F(t,t)=t,\qquad
\left.\partial_xF(x,s)\right|_{x=s=t}=\frac12,\qquad
\left.\partial_x\partial_sF(x,s)\right|_{x=s=t}=\frac1{2t}.
\label{eq:on-frequency-nonuniformity}
\end{equation}
The derivative limits are therefore not uniform at the origin.

To formulate the separate regulator question, insert a smooth cutoff
$\chi_{\rm null}(u/T)$ in each null integral, with
$\chi_{\rm null}=1$ near zero and compact support.  Let
$C_{\epsilon,T}(x,s)$ denote a matrix element of the commutator at
frequencies $(x,-s)$, with the two sheets separated by $\epsilon$ and
compact transverse smearings understood.
A moment-first calculation would use the sequence
\begin{equation}
\begin{aligned}
\text{form }C_{\epsilon,T}
&\;\longrightarrow\;\mathcal M_{ab}C_{\epsilon,T}
\;\longrightarrow\;T\to\infty
\\
&\;\longrightarrow\;\epsilon\to0^+
\;\longrightarrow\;\text{expand in }N^{-1}.
\end{aligned}
\label{eq:reg-order-of-operations}
\end{equation}
By contrast, eq.~\eqref{eq:on-frequency-commutator} gives the detector
limit at fixed frequencies.  Inferring the moment-first result from that
kernel would require
\begin{equation}
\lim_{\epsilon\to0^+}\lim_{T\to\infty}
\mathcal M_{ab}C_{\epsilon,T}
\ \overset{?}{=}\
\mathcal M_{ab}\!
\left(\lim_{\epsilon\to0^+}\lim_{T\to\infty}C_{\epsilon,T}\right).
\label{eq:on-unproved-moment-interchange}
\end{equation}
Equation~\eqref{eq:on-unproved-moment-interchange} formulates the remaining
analytic question.  Establishing this interchange requires uniform
estimates for the full regulated matrix elements, including their
frequency derivatives and contact terms.  The sheet separation is retained
while taking the light transforms \cite{Korchemsky:2021htm}.

For the displayed central term, replace
$\delta(\omega_1+\omega_2)$ by a smooth finite-time approximation
$\delta_T$.  At every finite $T$, the value and all derivatives with at
most one derivative in each frequency of
$\omega_1^3\delta_T(\omega_1+\omega_2)$ vanish at the origin.
This checks the stated central distribution under the chosen smoothing.

An isolated pairwise detector limit and a nested commutator are also
different operations.  At fixed frequencies, an operator of twist
$\tau>1$ contributes $\epsilon^{(\tau-1)/2}$ to the first commutator.
Its commutator with a third energy-flow operator can scale as
$\epsilon^{(1-\tau)/2}$, leaving a finite product in the regulated Jacobi
relation \cite{Korchemsky:2021htm}.

\section{Projected adjacent-mode conditions}
\label{app:projected-null-conditions}
\label{sec:lambda-null-classification}

This appendix derives the OPE actions and tests the selected adjacent set
\begin{equation}
\mathcal S_{\rm adj}
=\left\{w^p_{b,p-1}:\ 1-p\leq b\leq p-1\right\}
\cup\left\{w^p_{p-1,p-2}\right\},
\qquad 2p\in\mathbb Z_{\geq3},\quad b\in p+\mathbb Z.
\label{eq:null-selected-adjacent-set}
\end{equation}
The first set contains all adjacent double-pole modes; the second contains
only the top antiholomorphic component of the adjacent single-pole modes.
Thus $\mathcal S_{\rm adj}$ excludes
$P_{-,-}=w^{3/2}_{-1/2,-1/2}$ and the other single-pole components.
We test vanishing mode by mode after a Lorentz-global projection.
The translation obstruction to this projection is derived below.
The exact null descendants of
Section~\ref{sec:lambda-exact-null-descendants} instead belong to the
unprojected celestial representation.

\subsection{Residues and projected dimensions}
Writing $\mathcal G_i^+=G^+_{\Delta_i}(z_i,\bar z_i)$ for insertion labels,
set
$z_{ij}=z_i-z_j$ and $\bar z_{ij}=\bar z_i-\bar z_j$.  We use
$B(x,y)=\Gamma(x)\Gamma(y)/\Gamma(x+y)$ for the Euler beta function, and
$\kappa$ denotes the gravitational coupling in the normalization of
ref.~\cite{Taylor:2023ajb}.  A label $(4)$ means evaluation at
$(z_4,\bar z_4)$, with $\partial=\partial_{z_4}$ and
$\bar\partial=\partial_{\bar z_4}$.  The two curvature-dependent terms in the
positive-helicity graviton OPE are
\cite{Taylor:2023ajb}
\begin{align}
\mathcal G^+_3\mathcal G^+_4\big|_{\Lambda}
={}&\frac{\kappa\Lambda}{2z_{34}^{2}}(\Delta_3+\Delta_4)
\sum_{s\geq0}B(\Delta_3-2+s,\Delta_4-2)
\frac{\bar z_{34}^{s}}{s!}
\bar\partial^sG^+_{\Delta_3+\Delta_4-2}(4)
\nonumber\\
&+\frac{\kappa\Lambda}{2z_{34}}\Delta_3
\sum_{s\geq0}B(\Delta_3-2+s,\Delta_4-2)
\frac{\bar z_{34}^{s}}{s!}
\partial\bar\partial^sG^+_{\Delta_3+\Delta_4-2}(4).
\label{eq:null-curved-OPE}
\end{align}
The factors $\Delta_3+\Delta_4$ and $\Delta_3$ must be retained separately.
For the conformally soft residue
\begin{equation}
\Delta_3=4-2p+\epsilon,
\qquad
\Delta_4=\Delta,
\label{eq:null-soft-identification}
\end{equation}
where $\epsilon\to0$ is the residue parameter, they become
$\Delta+4-2p$ and $4-2p$, respectively.  We use the index
ordering of eq.~\eqref{eq:Lambda-w-bracket}.  Let $H^k_{a,b}$ denote the
$(a,b)$ mode of the conformally soft graviton current $H^k$ of
ref.~\cite{Taylor:2023ajb}.  This convention is fixed by
\begin{equation}
w^p_{b,a}
=\frac{1}{\kappa}\Gamma(p-b)\Gamma(p+b)H^{4-2p}_{a,b},
\qquad 2p\in\mathbb Z_{\geq3},\quad a,b\in p+\mathbb Z,
\quad 1-p\leq b\leq p-1,
\label{eq:null-soft-w-map}
\end{equation}
where $a$ is the unrestricted holomorphic mode on the indicated lattice.
Thus no additional
normalization is introduced in passing from the OPE to the algebra.

Define
\begin{equation}
F_{p,b}=\Gamma(p-b)\Gamma(p+b),
\quad
r=p+b-1,
\quad
n=p-1-b,
\quad
\mathcal R_r(\beta)
=\frac{(-1)^r}{r!}\frac{\Gamma(\beta)}{\Gamma(\beta-r)}.
\label{eq:null-residue-data}
\end{equation}
For nonnegative integer $r$, the ratio
$\Gamma(\beta)/\Gamma(\beta-r)=\prod_{j=1}^{r}(\beta-j)$ is understood
by polynomial continuation, including at poles of the separate gamma factors.
We use radial-quantization notation $\lvert G^+_\delta\rangle$ for the state
created by $G^+_\delta$ at the origin.
The beta-function residue
$\lim_{\epsilon\to0}\epsilon B(-r+\epsilon,\beta)
=\mathcal R_r(\beta)$ then gives the adjacent double-pole action
\begin{equation}
w^p_{b,p-1}\lvert G^+_{\Delta}\rangle
=\frac{\Lambda F_{p,b}}{2}(\Delta+4-2p)
\frac{\mathcal R_r(\Delta-2)}{n!}
\bar\partial^n\lvert G^+_{\Delta+2-2p}\rangle.
\label{eq:null-double-action}
\end{equation}
For the accompanying single-pole action, define
\begin{equation}
n_0=p-2-b,
\qquad
n_1=p-1-b.
\label{eq:null-app-mode-indices}
\end{equation}
Taking the residue of both the undeformed and curvature-dependent simple
poles gives
\begin{align}
w^p_{b,p-2}\lvert G^+_\Delta\rangle
={}&-\frac{F_{p,b}}{2}
\frac{\mathcal R_r(\Delta-1)}{n_0!}
\bar\partial^{n_0}\lvert G^+_{\Delta+4-2p}\rangle
\nonumber\\
&+\frac{\Lambda F_{p,b}(4-2p)}{2}
\frac{\mathcal R_r(\Delta-2)}{n_1!}
\partial\bar\partial^{n_1}
\lvert G^+_{\Delta+2-2p}\rangle,
\label{eq:null-app-single-action}
\end{align}
where a term with negative $n_0$ or $n_1$ is absent.  For $b=p-1$, the
undeformed contribution is absent and the curvature term reduces to
\begin{equation}
w^p_{p-1,p-2}\lvert G^+_\Delta\rangle
=\frac{\Lambda F_{p,p-1}(4-2p)}{2}
\mathcal R_{2p-2}(\Delta-2)
\partial\lvert G^+_{\Delta+2-2p}\rangle.
\label{eq:null-app-top-single-action}
\end{equation}
Equations~\eqref{eq:null-double-action} and
\eqref{eq:null-app-top-single-action} are the conditions on
$\mathcal S_{\rm adj}$.  Equation~\eqref{eq:null-app-single-action}
also records the other single-pole actions, but their vanishing is not
imposed.  Modes with holomorphic index greater than $p-1$ already
annihilate the hard primary.

For a positive-helicity primary of dimension $\delta$, the first
antiholomorphic global null level is
\begin{equation}
N_{\rm null}=1-2\bar h=3-\delta,
\label{eq:null-global-level}
\end{equation}
whenever the right-hand side is a positive integer.  At $\Delta=3$, the
state on the right-hand side of eq.~\eqref{eq:null-double-action} has
dimension $\delta=5-2p$.  Since
\begin{equation}
r+n=2p-2=N_{\rm null},
\label{eq:null-rn-identity}
\end{equation}
every double-pole image in $\mathcal S_{\rm adj}$ vanishes after
the global projection.  If
$r\geq1$, its coefficient contains
$\mathcal R_r(1)=0$; if $r=0$, then $n=2p-2$ and the descendant itself is at
the first null level.  The top single-pole image
\eqref{eq:null-app-top-single-action} vanishes by the same residue
mechanism.  Consequently,
\begin{equation}
\Delta=3
\label{eq:null-delta3-projected}
\end{equation}
is the unique positive-dimension point at which all images of the
selected set $\mathcal S_{\rm adj}$ vanish under this projection.

The $p=2$ double-pole family makes the additional candidate $\Delta=4$
explicit:
\begin{align}
w^2_{-1,1}\lvert G^+_\Delta\rangle
&=\frac{\Lambda\Delta}{2}
\bar\partial^2\lvert G^+_{\Delta-2}\rangle,
\nonumber\\
w^2_{0,1}\lvert G^+_\Delta\rangle
&=-\frac{\Lambda\Delta(\Delta-3)}{2}
\bar\partial\lvert G^+_{\Delta-2}\rangle,
\nonumber\\
w^2_{1,1}\lvert G^+_\Delta\rangle
&=\frac{\Lambda\Delta(\Delta-3)(\Delta-4)}{2}
\lvert G^+_{\Delta-2}\rangle.
\label{eq:null-app-p2-actions}
\end{align}
The $p=3/2$ translation modes remove this candidate:
\begin{align}
w^{3/2}_{-1/2,1/2}\lvert G^+_4\rangle
&=\frac{5\Lambda}{2}\bar\partial\lvert G^+_3\rangle,
\nonumber\\
w^{3/2}_{1/2,1/2}\lvert G^+_4\rangle
&=-\frac{5\Lambda}{2}\lvert G^+_3\rangle,
\nonumber\\
w^{3/2}_{1/2,-1/2}\lvert G^+_4\rangle
&=-\frac{\Lambda}{2}\partial\lvert G^+_3\rangle.
\label{eq:null-app-delta4-translations}
\end{align}
For a positive-helicity primary $G^+_\delta$ with
$h=(\delta+2)/2$ and $\bar h=(\delta-2)/2$, the Shapovalov pairing at
$m,n\in\mathbb Z_{\geq0}$ is
\begin{equation}
\frac{\left\|L_{-1}^{m}\bar L_{-1}^{n}
G^+_\delta\right\|^2}{\left\|G^+_\delta\right\|^2}
=m!(2h)_m\,n!(2\bar h)_n.
\label{eq:null-app-global-norm}
\end{equation}
Here $(x)_m=\Gamma(x+m)/\Gamma(x)$ is the rising Pochhammer symbol.
For $m\in\mathbb Z_{\geq0}$, its polynomial continuation is
$(x)_m=\prod_{j=0}^{m-1}(x+j)$, with $(x)_0=1$; this definition also
applies at nonpositive integer $x$.
The pairing vanishes on $\bar\partial^2G^+_1$,
$\bar\partial^2G^+_2$, and $\bar\partial G^+_2$.  By contrast, for real
$\Lambda$ the final state in
eq.~\eqref{eq:null-app-delta4-translations} obeys
\begin{equation}
\left\|w^{3/2}_{1/2,-1/2}G^+_4\right\|^2
=\frac{5\Lambda^2}{4}\left\|G^+_3\right\|^2.
\label{eq:null-delta4-representative}
\end{equation}
Hence $\Delta=4$ passes the projected $p=2$ conditions, but not all
conditions in $\mathcal S_{\rm adj}$.

The preceding examples can be promoted to a classification that also keeps
negative and nonstandard Mellin dimensions.  The three $p=3/2$ images at
generic $\Delta$ are
\begin{align}
w^{3/2}_{-1/2,1/2}\lvert G^+_\Delta\rangle
&=\frac{\Lambda}{2}(\Delta+1)
\bar\partial\lvert G^+_{\Delta-1}\rangle,
\nonumber\\
w^{3/2}_{1/2,1/2}\lvert G^+_\Delta\rangle
&=-\frac{\Lambda}{2}(\Delta-3)(\Delta+1)
\lvert G^+_{\Delta-1}\rangle,
\nonumber\\
w^{3/2}_{1/2,-1/2}\lvert G^+_\Delta\rangle
&=-\frac{\Lambda}{2}(\Delta-3)
\partial\lvert G^+_{\Delta-1}\rangle.
\label{eq:null-generic-translations}
\end{align}
Denote the remaining $p=3/2$ mode by
$P_{-,-}=w^{3/2}_{-1/2,-1/2}$.  Its action is
\begin{equation}
P_{-,-}\lvert G^+_\Delta\rangle
=-\frac12\lvert G^+_{\Delta+1}\rangle
+\frac{\Lambda}{2}\partial\bar\partial
 \lvert G^+_{\Delta-1}\rangle.
\label{eq:null-fourth-translation-action}
\end{equation}
The three curvature terms in eq.~\eqref{eq:null-generic-translations}
lower the Mellin dimension.  The flat-space momentum term in $P_{-,-}$
raises it.
The middle image is a primary.  For nonzero $\Lambda$, its coefficient
restricts a projected solution to $\Delta=-1$ or $\Delta=3$.  At both values,
the remaining translation images vanish after projection by global nulls.
At $\Delta=-1$, the final image in
eq.~\eqref{eq:null-generic-translations} is proportional to
$\partial G^+_{-2}$, whose first holomorphic null level is one.

The $p=2$ family removes this exceptional negative value.  Its $b=-1$
component gives
\begin{equation}
w^2_{-1,1}\lvert G^+_{-1}\rangle
=-\frac{\Lambda}{2}\bar\partial^2
\lvert G^+_{-3}\rangle.
\label{eq:null-minus-one-exclusion}
\end{equation}
The first antiholomorphic null of $G^+_{-3}$ occurs at level six, so the
level-two state in eq.~\eqref{eq:null-minus-one-exclusion} is nonzero after
global projection.  The complete classification of these projected
low-family conditions is
summarized in Table~\ref{tab:null-low-family-classification}.
\begin{table}[tbp]
\centering
\begin{tabular}{c|c|c}
modes imposed & common coefficient zeros & after global projection
\\ \hline
$p=2$ & $0$ & $0,3,4$ \\
$p=3/2$ & $\varnothing$ & $-1,3$ \\
$p=3/2,2$ & $\varnothing$ & $3$ \\
$\mathcal S_{\rm adj}$ with $p\leq3$ & $\varnothing$ & $3$
\end{tabular}
\caption{Exceptional dimensions for the selected conditions
$\mathcal S_{\rm adj}$ in the displayed low-mode families.  The middle
column lists common coefficient zeros before global projection.
The remaining single-pole modes are not imposed.}
\label{tab:null-low-family-classification}
\end{table}
The primary image in eq.~\eqref{eq:null-generic-translations} makes the
second row exact rather than a finite scan.  Combining it with
eq.~\eqref{eq:null-minus-one-exclusion} leaves only $\Delta=3$.  The residue
argument following eq.~\eqref{eq:null-rn-identity} then proves that every
image from $\mathcal S_{\rm adj}$ at this value vanishes after global
projection for $p\geq3/2$.

\subsection{The translation obstruction}

The opposite pair among the projected translation conditions is
\begin{equation}
P_{+,-}=w^{3/2}_{-1/2,1/2},
\qquad
P_{-,+}=w^{3/2}_{1/2,-1/2}.
\label{eq:null-opposite-translations}
\end{equation}
Its commutator is
\begin{equation}
[P_{+,-},P_{-,+}]
=\frac{\Lambda}{2}(\bar L_0-L_0).
\label{eq:null-opposite-commutator}
\end{equation}
At $\Delta=3$, the actions are
\begin{equation}
P_{+,-}\lvert G_3^+\rangle
=2\Lambda\bar\partial\lvert G_2^+\rangle,
\qquad
P_{-,+}\lvert G_3^+\rangle=0.
\label{eq:null-delta3-opposite-actions}
\end{equation}
Since $G_3^+$ has weights $(5/2,1/2)$,
\begin{equation}
-2\Lambda P_{-,+}\bar\partial\lvert G_2^+\rangle
=[P_{+,-},P_{-,+}]\lvert G_3^+\rangle
=-\Lambda\lvert G_3^+\rangle.
\label{eq:null-return-map-derivation}
\end{equation}
For $\Lambda\ne0$ this gives
eq.~\eqref{eq:null-global-quotient-failure}.  The same cancellation is valid
in a $\Lambda$-torsion-free formal module, where $\Lambda v=0$ implies
$v=0$.  The discarded global-null subspace is therefore not invariant:
quotienting by $\bar\partial G_2^+$ would also remove $G_3^+$.

The alternative closed set
$\mathfrak n_L=\operatorname{span}\{L_1,\bar L_1,P_{+,-},P_{+,+}\}$
does not annihilate $G_3^+$ before projection.

\section{Low-spin module and Ward-operator checks}
\label{app:celestial-module-checks}

\subsection{Descendant action and the Casimir}

To check the covariance of the unrestricted low-spin module, define the
descendant states
\begin{equation}
\lvert G^+_\delta;m,n\rangle
=L_{-1}^m\bar L_{-1}^n\lvert G^+_\delta\rangle,
\qquad m,n\in\mathbb Z_{\geq0}.
\label{eq:null-low-pbw-states}
\end{equation}
The hard-state residue convention gives
$L_{-1}G^+_\delta=\partial G^+_\delta$ and
$\bar L_{-1}G^+_\delta=-\bar\partial G^+_\delta$.  The mixed brackets then
determine the translation action recursively.  For $m>0$,
\begin{align}
P_{a,b}\lvert G^+_\delta;m,n\rangle
={}&L_{-1}P_{a,b}\lvert G^+_\delta;m-1,n\rangle
+\left(a+\frac12\right)
 P_{a-1,b}\lvert G^+_\delta;m-1,n\rangle,
\label{eq:null-low-hol-recursion}
\end{align}
while for $m=0$ and $n>0$,
\begin{align}
P_{a,b}\lvert G^+_\delta;0,n\rangle
={}&\bar L_{-1}P_{a,b}\lvert G^+_\delta;0,n-1\rangle
+\left(b+\frac12\right)
 P_{a,b-1}\lvert G^+_\delta;0,n-1\rangle.
\label{eq:null-low-anti-recursion}
\end{align}
Solving these recursions gives the closed all-level actions
\begin{align}
P_{-,-}\lvert G_\delta^+;m,n\rangle
={}&-\frac12\lvert G_{\delta+1}^+;m,n\rangle
-\frac{\Lambda}{2}\lvert G_{\delta-1}^+;m+1,n+1\rangle,
\nonumber\\
P_{-,+}\lvert G_\delta^+;m,n\rangle
={}&-\frac n2\lvert G_{\delta+1}^+;m,n-1\rangle
-\frac{\Lambda}{2}(\delta+n-3)
\lvert G_{\delta-1}^+;m+1,n\rangle,
\nonumber\\
P_{+,-}\lvert G_\delta^+;m,n\rangle
={}&-\frac m2\lvert G_{\delta+1}^+;m-1,n\rangle
-\frac{\Lambda}{2}(\delta+m+1)
\lvert G_{\delta-1}^+;m,n+1\rangle,
\nonumber\\
P_{+,+}\lvert G_\delta^+;m,n\rangle
={}&-\frac{mn}{2}\lvert G_{\delta+1}^+;m-1,n-1\rangle
\nonumber\\
&-\frac{\Lambda}{2}(\delta+m+1)(\delta+n-3)
\lvert G_{\delta-1}^+;m,n\rangle.
\label{eq:null-all-level-translations}
\end{align}
Terms with negative descendant indices have zero coefficients and are
omitted.  For example, commuting $P_{+,-}$ past $L_{-1}^m$ adds
$mL_{-1}^{m-1}P_{-,-}$; its downward-shift term supplies the $m$ in
$\delta+m+1$.  Both recursions and the primary actions then prove
eq.~\eqref{eq:null-all-level-translations} by induction.

For completeness, with $h=(\delta+2)/2$ and $\bar h=(\delta-2)/2$,
the nontrivial Lorentz actions are
\begin{align}
L_0\lvert G_\delta^+;m,n\rangle
&=(h+m)\lvert G_\delta^+;m,n\rangle,
\nonumber\\
\bar L_0\lvert G_\delta^+;m,n\rangle
&=(\bar h+n)\lvert G_\delta^+;m,n\rangle,
\nonumber\\
L_1\lvert G_\delta^+;m,n\rangle
&=m(\delta+m+1)\lvert G_\delta^+;m-1,n\rangle,
\nonumber\\
\bar L_1\lvert G_\delta^+;m,n\rangle
&=n(\delta+n-3)\lvert G_\delta^+;m,n-1\rangle.
\label{eq:null-all-level-lorentz}
\end{align}
The two lowering modes raise the corresponding descendant index by one.
Using these formulas, all $45$ low-spin generator-pair relations hold
as polynomial identities with $m,n,\delta,\Lambda$ symbolic.
The same formulas reproduce eq.~\eqref{eq:null-global-quotient-failure}
with the derivative convention above and give
$\mathcal C_2\lvert G_\delta^+;m,n\rangle
=3\lvert G_\delta^+;m,n\rangle$ at arbitrary levels.

For $\Lambda\ne0$, define the Killing form by
$K(X,Y)=\operatorname{Tr}_{\rm ad}(\operatorname{ad}_X\operatorname{ad}_Y)$.
In the ordered basis
\[
(L_{-1},L_0,L_1,\bar L_{-1},\bar L_0,\bar L_1,
P_{-,-},P_{-,+},P_{+,-},P_{+,+}),
\]
its independent nonzero entries are
\begin{align}
K(L_{-1},L_1)&=-6,
&
K(L_0,L_0)&=3,
\nonumber\\
K(\bar L_{-1},\bar L_1)&=-6,
&
K(\bar L_0,\bar L_0)&=3,
\nonumber\\
K(P_{-,-},P_{+,+})&=-3\Lambda,
&
K(P_{-,+},P_{+,-})&=3\Lambda,
\label{eq:null-killing-form}
\end{align}
together with their symmetric counterparts.  Three times the inverse-Killing
Casimir is eq.~\eqref{eq:null-quadratic-casimir}; invariance of the form
implies its centrality.

Using the descendant recursions, its two translation anticommutators give
\begin{align}
\frac1\Lambda\{P_{-,+},P_{+,-}\}\lvert G^+_\Delta\rangle
={}&\frac{\Delta-1}{2}\lvert G^+_\Delta\rangle
+\frac{\Lambda}{2}(\Delta-3)(\Delta+1)
 \lvert G^+_{\Delta-2};1,1\rangle,
\\
-\frac1\Lambda\{P_{-,-},P_{+,+}\}\lvert G^+_\Delta\rangle
={}&\frac{3+\Delta-\Delta^2}{2}\lvert G^+_\Delta\rangle
-\frac{\Lambda}{2}(\Delta-3)(\Delta+1)
 \lvert G^+_{\Delta-2};1,1\rangle.
\label{eq:null-translation-anticommutators}
\end{align}
Their shifted descendants cancel.  The translation contribution is
\begin{equation}
\frac{-\Delta^2+2\Delta+2}{2}\lvert G^+_\Delta\rangle.
\label{eq:null-translation-casimir-piece}
\end{equation}
and the Lorentz contribution is
\begin{equation}
\left[h(h-1)+\bar h(\bar h-1)\right]\lvert G^+_\Delta\rangle
=\frac{\Delta^2-2\Delta+4}{2}\lvert G^+_\Delta\rangle.
\label{eq:null-lorentz-casimir-piece}
\end{equation}
Their sum yields $\mathcal C_2=3$ on each hard primary.
For a general low-spin highest-weight vector,
\begin{equation}
L_1v=\bar L_1v=P_{-,+}v=P_{+,+}v=0,
\qquad L_0v=hv,\qquad \bar L_0v=\bar h v,
\label{eq:null-low-spin-highest-weight}
\end{equation}
the translation line of the Casimir instead contributes $-\bar h$, giving
\begin{equation}
\begin{aligned}
\mathcal C_2v
&=\left[h(h-1)+\bar h(\bar h-1)-\bar h\right]v
\\
&=\frac12\left[E_0^{\rm AdS}(E_0^{\rm AdS}-3)+s(s+1)\right]v,
\qquad E_0^{\rm AdS}=h+\bar h,\quad s=h-\bar h.
\end{aligned}
\label{eq:null-low-spin-highest-weight-casimir}
\end{equation}
Here $E_0^{\rm AdS}=h+\bar h$ and $s=h-\bar h$ are the highest-weight energy and spin
labels used in the AdS comparison.

\subsection{Singular-vector checks and the AdS norm}
\label{app:celestial-shortening-checks}

The cubic polynomial has the following action at general Mellin dimension:
\begin{align}
&\left(L_{-1}P_{+,-}^2-3P_{-,-}P_{+,-}
+3\Lambda\bar L_{-1}\right)\lvert G_\delta^+\rangle
\nonumber\\
&\qquad=(\delta-3)\left[
\frac{\Lambda^2}{4}(\delta+1)L_{-1}\bar L_{-1}^2
\lvert G_{\delta-2}^+\rangle
-\frac{3\Lambda}{4}\bar L_{-1}\lvert G_\delta^+\rangle\right].
\label{eq:null-generic-delta-cubic}
\end{align}
This verifies the zero at $\delta=3$ without removing either basis
descendant.  The first four steps of the other null relation are
\begin{align}
P_{+,-}\lvert G_3^+\rangle
&=-2\Lambda\bar L_{-1}\lvert G_2^+\rangle,
&
P_{+,-}^2\lvert G_3^+\rangle
&=3\Lambda^2\bar L_{-1}^2\lvert G_1^+\rangle,
\nonumber\\
P_{+,-}^3\lvert G_3^+\rangle
&=-3\Lambda^3\bar L_{-1}^3\lvert G_0^+\rangle,
&
P_{+,-}^4\lvert G_3^+\rangle
&=\frac32\Lambda^4\bar L_{-1}^4\lvert G_{-1}^+\rangle.
\label{eq:null-first-four-powers}
\end{align}

For the energy-dependent shortening calculation, let
$\lvert\Omega_E\rangle$ have weights
$((E+2)/2,(E-2)/2)$ and be annihilated by $\mathfrak n_R$.
Define $\lvert\chi_E\rangle$ by the same polynomial as in
eq.~\eqref{eq:null-chi-singular}.  PBW reduction gives
\begin{align}
P_{-,+}\lvert\chi_E\rangle&=0,
&
L_1\lvert\chi_E\rangle&=(E-3)P_{+,-}^2\lvert\Omega_E\rangle,
\nonumber\\
\bar L_1\lvert\chi_E\rangle&=3\Lambda(E-3)\lvert\Omega_E\rangle,
&
P_{+,+}\lvert\chi_E\rangle
&=-\frac{3\Lambda}{2}(E-3)P_{+,-}\lvert\Omega_E\rangle.
\label{eq:null-energy-dependent-raising}
\end{align}
Thus the spin-one descendant becomes singular at $E=3$.
The identity
$P_{+,-}^3\lvert\chi_E\rangle
=L_{-1}P_{+,-}^5\lvert\Omega_E\rangle$
also holds before specializing $E$.

For the norm calculation only, choose the AdS positive-energy adjoint
with real $\Lambda\ne0$,
\begin{align}
L_{-1}^\dagger&=L_1,
&
P_{+,-}^\dagger&=\frac4\Lambda P_{-,+},
\nonumber\\
P_{-,-}^\dagger&=\frac4\Lambda P_{+,+},
&
\bar L_{-1}^\dagger&=\frac{16}{\Lambda^2}\bar L_1,
\label{eq:null-ads-adjoint}
\end{align}
and $L_0^\dagger=L_0$, $\bar L_0^\dagger=\bar L_0$.
It respects all low-spin brackets.  Commuting the adjoint word to the
right gives the contravariant norm
\begin{equation}
\langle\chi_E|\chi_E\rangle
=240(E-3)\langle\Omega_E|\Omega_E\rangle.
\label{eq:null-chi-norm}
\end{equation}
The inverse powers of $\Lambda$ enter the adjoint normalization only,
not the state identities or Ward equations.

The map to Dolan's generators is
\begin{equation}
(P_{1+},P_0,P_{1-},E_1^-)_{\rm Dolan}
=\left(2L_{-1},-\ii P_{-,-},
\frac{\Lambda}{2}\bar L_{-1},P_{+,-}\right).
\label{eq:null-dolan-map}
\end{equation}
At spin two, his Eq.~(C.5) is $2\lvert\chi_E\rangle$, and Eq.~(C.6)
has four times the norm in \eqref{eq:null-chi-norm}
\cite{Dolan:2005wy}.  This identifies the level-one shortening, in addition
to the common Casimir value of the $(3,2)$ and $(4,1)$ modules.
The zero norm occurs in the reducible module; the positive-definite
irreducible quotient removes its null submodule.

\subsection{Ward-operator ordering checks}
\label{app:celestial-ward-ordering}

The one-leg square of the transported operator is
\begin{align}
&\left(d_{+,-}^{(i)}+z_{i0}d_{-,-}^{(i)}\right)^2
=\frac{z_{i0}^2}{4}(\mathsf E_i^+)^2
-\frac{\Lambda z_{i0}}{2}(\Delta_i+2+z_{i0}\partial_i)\bar\partial_i
\nonumber\\
&\qquad\quad+\frac{\Lambda^2}{4}
(\Delta_i+1+z_{i0}\partial_i)(\Delta_i+z_{i0}\partial_i)
\bar\partial_i^2(\mathsf E_i^-)^2.
\label{eq:null-transported-square}
\end{align}
The middle coefficient $\Delta_i+2$ receives contributions from both the
Mellin shift and differentiation of $z_{i0}$.  This expression is checked
by composition on an arbitrary symbolic function.

The external operator commutators are
\begin{align}
\left[\sum_i\partial_i,\,
\sum_j(d_{+,-}^{(j)}+z_{j0}d_{-,-}^{(j)})\right]
&=\sum_i d_{-,-}^{(i)},
\nonumber\\
\left[\sum_i d_{-,-}^{(i)},\,
\sum_j(d_{+,-}^{(j)}+z_{j0}d_{-,-}^{(j)})\right]
&=-\frac{\Lambda}{2}\sum_i\bar\partial_i .
\label{eq:null-ward-kernel-brackets}
\end{align}
Using them to place the holomorphic derivatives on the right gives
\begin{align}
\mathcal C_\chi={}&
\left(\sum_i(d_{+,-}^{(i)}+z_{i0}d_{-,-}^{(i)})\right)^2
\sum_j\partial_j
\nonumber\\
&+5\sum_{i,j}(d_{+,-}^{(i)}+z_{i0}d_{-,-}^{(i)})d_{-,-}^{(j)}
-5\Lambda\sum_i\bar\partial_i .
\label{eq:null-right-ordered}
\end{align}
The coefficients in this order are $5$ and $-5\Lambda$.
They differ from those in eq.~\eqref{eq:null-cubic-ward-operator}
because neither the derivatives nor the two translation actions commute
through the preceding coefficients.

A check at the marked insertion keeps its coordinate dependence explicit:
\begin{equation}
\partial_0\left(d_{+,-}^{(i)}+z_{i0}d_{-,-}^{(i)}\right)
=-d_{-,-}^{(i)}.
\label{eq:null-center-kernel-derivative}
\end{equation}
The derivative of the squared sum therefore contains both product orders.
Combining this identity with the global Ward identity
$\partial_0\mathcal A_{n+1}=-\sum_i\partial_i\mathcal A_{n+1}$ gives
\begin{align}
&\partial_0\left\{
\left(\sum_j(d_{+,-}^{(j)}+z_{j0}d_{-,-}^{(j)})\right)^2
\mathcal A_{n+1}\right\}
\nonumber\\
&\qquad=-\sum_i\partial_i
\left(\sum_j(d_{+,-}^{(j)}+z_{j0}d_{-,-}^{(j)})\right)^2
\mathcal A_{n+1}.
\label{eq:null-center-derivative}
\end{align}
This reproduces the first line of eq.~\eqref{eq:null-three-ward-images}.

\subsection{An exact one-leg reduction}
\label{app:celestial-ward-independence}

The two singular vectors specify additional relations in the concrete
celestial module.  External-leg identities can relate their Ward images
to the highest-weight and global Ward operators.  For one external leg,
this relation is explicit.

For one external leg, the transported Lorentz actions are
\begin{align}
d_{L_0}^{(1)}&=z_{10}\partial_1+\frac{\Delta_1+2}{2},
&d_{\bar L_0}^{(1)}&=\bar z_{10}\bar\partial_1+\frac{\Delta_1-2}{2},
\nonumber\\
d_{L_1}^{(1)}&=z_{10}^2\partial_1+(\Delta_1+2)z_{10},
&d_{\bar L_1}^{(1)}&=-\bar z_{10}^2\bar\partial_1-(\Delta_1-2)\bar z_{10}.
\label{eq:null-one-leg-lorentz}
\end{align}
The old Ward conditions include
\begin{align}
d_{L_1}^{(1)}\mathcal A_2&=d_{\bar L_1}^{(1)}\mathcal A_2=0,
\nonumber\\
\left(d_{L_0}^{(1)}+\frac52\right)\mathcal A_2&=0,
&\left(d_{\bar L_0}^{(1)}+\frac12\right)\mathcal A_2&=0,
\label{eq:null-one-leg-old-conformal}
\end{align}
together with eqs.~\eqref{eq:null-first-ward-equation} and
\eqref{eq:null-second-ward-equation}.  The plus signs in the Cartan
conditions come from moving the marked state's weights through the
Ward identity.  The cubic operator obeys the exact factorization
\begin{align}
\mathcal C_\chi={}&
-(d_{-,-}^{(1)})^2d_{L_1}^{(1)}-\Lambda\bar\partial_1^2d_{\bar L_1}^{(1)}
\nonumber\\
&-2(d_{+,-}^{(1)}+z_{10}d_{-,-}^{(1)})\bar\partial_1
(d_{-,+}^{(1)}-\bar z_{10}d_{-,-}^{(1)})
\nonumber\\
&+2\bar\partial_1d_{-,-}^{(1)}
\left(d_{+,+}^{(1)}+z_{10}d_{-,+}^{(1)}
-\bar z_{10}d_{+,-}^{(1)}-z_{10}\bar z_{10}d_{-,-}^{(1)}\right)
\nonumber\\
&+\left[2(d_{+,-}^{(1)}+z_{10}d_{-,-}^{(1)})d_{-,-}^{(1)}
-\frac{\Lambda}{2}\bar\partial_1\right]\left(d_{L_0}^{(1)}+\frac52\right)
\nonumber\\
&-\Lambda\bar\partial_1\left(d_{\bar L_0}^{(1)}+\frac32\right)
\left(d_{\bar L_0}^{(1)}+\frac12\right).
\label{eq:null-one-leg-certificate}
\end{align}
Every term has an old Ward condition as its rightmost factor.
This is an identity on an arbitrary one-leg function, without division
by $\Lambda$.  It proves that the cubic equation is redundant for one
external leg.

For two and three external legs, the directly summed one-leg expression
fails as an operator identity.  An exact search in the 22-candidate
curvature-homogeneous ansatz, formed from at most two global generators
multiplying the existing Ward conditions, finds no reduction for those
two cases.

\bibliographystyle{JHEP}
\bibliography{biblio}

\end{document}